\documentclass[apj,twocolumn]{emulateapj}

\usepackage{natbib}
\usepackage{amsmath}
\usepackage{graphics,color}

\def\mathbi#1{\textbf{\em #1}}
\def\fnl{f_\mathrm{NL}}
\def\gnl{g_\mathrm{NL}}
\def\Mr{\mathcal{M}_R}
\def\Fr{\mathcal{F}_R}
\def\Gr{\mathcal{G}_R}
\def\Mm{\mathcal{M}}
\def\Wr{\tilde{W}_R}
\def\M0{\mathcal{M}_0}

\shorttitle{Primordial non-Gaussianity and galaxy bispectrum}
\shortauthors{JEONG \& KOMATSU}
\begin{document}
\title{%
   Primordial non-Gaussianity, scale-dependent bias, and the bispectrum
   of galaxies
}%
\author{
Donghui Jeong and Eiichiro Komatsu
}
\affil{Texas Cosmology Center, University of Texas at Austin, \\ 
       1 University Station, C1400, Austin, TX, 78712, USA
}
\email{djeong@astro.as.utexas.edu}
\begin{abstract}
  The three-point correlation function of cosmological fluctuations is a
 sensitive probe of the physics of inflation. We calculate the
 bispectrum, $B_g(k_1,k_2,k_3)$, Fourier transform of the three-point function
 of density peaks (e.g., galaxies), using two different methods: the
 Matarrese-Lucchin-Bonometto formula and the locality of galaxy bias. 
 The bispectrum of peaks is not only  sensitive to that of the 
 underlying matter density fluctuations, but also to the
 four-point function. For a physically-motivated, local form of primordial
 non-Gaussianity in the curvature perturbation,
 $\Phi=\phi+\fnl\phi^2+\gnl\phi^3$, where $\phi$ is a Gaussian
 field, we show that the galaxy bispectrum contains five physically
 distinct pieces: (i) non-linear gravitational evolution, (ii)
 non-linear galaxy bias, (iii) $\fnl$, (iv) $\fnl^2$, and (v)
 $\gnl$. While (i), (ii), and a part of (iii) have been derived in the
 literature, (iv) and (v) are derived in this paper for the
 first time. We also find that (iii) receives an enhancement of a factor
 of $\sim 15$ relative to the previous calculation for the squeezed
 triangles  ($k_1\approx k_2\gg k_3$).
 Our finding suggests that the galaxy bispectrum is
 more sensitive to $\fnl$ than previously recognized, and is also
 sensitive to a new term, $\gnl$. For a more general form of local-type
 non-Gaussianity, the coefficient $\fnl^2$ can be interpreted as
 $\tau_{\rm NL}$, which allows us to test multi-field inflation models
 using the relation between the three- and four-point functions.
 The usual terms from Gaussian initial
 conditions, (i) and (ii), have the smallest signals in the squeezed
 configurations, while 
 the others have the largest  signals; thus, we can
 distinguish them  easily. We cannot 
 interpret the 
 effects of $\fnl$ on $B_g(k_1,k_2,k_3)$ as a
 scale-dependent bias, and thus replacing
 the linear bias in the galaxy bispectrum with the scale-dependent bias
 known for the power spectrum results in an incorrect prediction.
 As the importance of primordial non-Gaussianity
 relative to the non-linear gravity evolution and galaxy bias
 increases toward higher redshifts, galaxy surveys probing a
 high-redshift universe are particularly useful for
 probing the primordial non-Gaussianity.
\end{abstract}
\section{Introduction}
Are primordial fluctuations Gaussian, or non-Gaussian? The simplest
models of inflation, driven by a slowly-rolling single scalar field
with the canonical kinetic term originated from the Bunch-Davis vacuum,
predict the amplitude of primordial non-Gaussianity that is below the
detectable level. Therefore, a convincing detection of primordial
non-Gaussianity 
would rule out the above simplest models, and thus lead to 
a breakthrough in our understanding of the physics of inflation
\citep[see][for a review]{bartolo/etal:2004}.

The tightest limit on primordial
non-Gaussianity so far comes from the angular bispectrum,
spherical harmonic transform of the angular three-point correlation
function \citep[see][for a review]{komatsu:prep},
 of anisotropy in the cosmic microwave background (CMB) radiation
\citep[see][for the latest limits]{komatsu/etal:2009,smith/senatore/zaldarriaga:prep,curto/martinez-gonzalez/barreiro:prep}.

The large-scale structure of the universe can also provide 
alternative ways of probing primordial non-Gaussianity through abundances 
and clustering properties of galaxies and clusters of galaxies. 
However, as the large-scale structure of the universe is more non-linear
than CMB, it was generally thought that CMB would be the most promising way of
constraining primordial non-Gaussianity \citep{verde/etal:2000}.

On the other hand, \citet{sefusatti/komatsu:2007} have shown that
observations of the large-scale structure of the universe in a {\it
high-redshift universe}, i.e., $z>1$, can provide competitive limits on
primordial non-Gaussianity, as the other non-linear effects are weaker
in a high redshift universe. 
Specifically, they calculate the bispectrum of the three-dimensional
distribution of galaxies, $B_g(k_1,k_2,k_3)$,\footnote{The bispectrum,
the Fourier transform of the three-point correlation function, 
is defined as
$\langle
\delta(\mathbi{k}_1)
\delta(\mathbi{k}_2)
\delta(\mathbi{k}_3)
\rangle
\equiv
(2\pi)^3 B(k_1,k_2,k_3)
\delta^D(\mathbi{k}_1+\mathbi{k}_2+\mathbi{k}_3)$.} on large scales 
as (see also \citet{scoccimarro/etal:2004})
\begin{eqnarray}
\nonumber
& & B_g(k_1,k_2,k_3,z)\\
\nonumber
&=& 3b_1^3\fnl\Omega_mH_0^2\left[\frac{P_m(k_1,z)}{k_1^2T(k_1)}
\frac{P_m(k_2,z)}{k_2^2T(k_2)}\frac{k_3^2T(k_3)}{D(z)}+(2~{\rm cyclic})\right]\\
\nonumber
&+& 2b_1^3\left[F_2^{(s)}({\mathbi k}_1,{\mathbi
		     k}_2)P_m(k_1,z)P_m(k_2,z)+(2~{\rm cyclic})\right]\\
&+& b_1^2b_2\left[P_m(k_1,z)P_m(k_2,z)+(2~{\rm cyclic})\right],
\label{eq:SK07}
\end{eqnarray}
where  $H_0$ and $\Omega_m$ are the present-day value of Hubble's
constant and the matter density parameter, respectively, 
$P_m(k,z)$ is the  power spectrum of linear matter density
fluctuations, $D(z)$ is the linear growth
factor, 
$T(k)$ is the linear transfer function whose limit is $T(k)\to 1$ as
$k\to 0$, and $F_2^{(s)}({\mathbi k}_1,{\mathbi k}_2)$ is a known mathematical
function given by \citep{bernardeau/etal:2002}
\begin{equation}
F_2^{(s)}(\mathbi{k}_1,\mathbi{k}_2)
=\frac{5}{7}
+\frac{\mathbi{k}_1\cdot\mathbi{k}_2}{2k_1k_2}
\left(
\frac{k_1}{k_2}+\frac{k_2}{k_1}
\right)
+\frac{2}{7}\left(\frac{\mathbi{k}_1\cdot\mathbi{k}_2}{k_1k_2}\right)^2.
\label{eq:F2}
\end{equation}
This function vanishes in the squeezed limit,
$\mathbi{k}_1=-\mathbi{k}_2$  (the triangles with the
maximum angle, i.e., $\pi$, between $\mathbi{k}_1$ and $\mathbi{k}_2$,
and $|\mathbi{k}_1|=|\mathbi{k}_2|$), and takes on
the maximum value, 
$F_2^{(s)}=(\alpha+1)^2/(2\alpha)\ge 2$, in the opposite limit,
$\mathbi{k}_1=\alpha\mathbi{k}_2$ where $\alpha\ge 1$ 
(the triangles with the 
vanishing angle between $\mathbi{k}_1$ and $\mathbi{k}_2$).

Here, 
$b_1$ and $b_2$ are the linear and non-linear galaxy bias parameters,
respectively, which relate the underlying matter density contrast,
$\delta_m$, to the galaxy density contrast, $\delta_g$, as
\citep{fry/gaztanaga:1993}
\begin{equation}\label{eq:def_local_bias}
 \delta_g({\mathbi x})=b_1\delta_m({\mathbi x}) 
+ \frac{b_2}2
\left[
\delta_m^2({\mathbi x}) - \sigma^2
\right]
+\cdots,
\end{equation}
where $\sigma^2\equiv \langle \delta_m^2\rangle$, which ensures $\langle\delta_g\rangle=0$.

The last two terms in Eq.~(\ref{eq:SK07}) are the well-known results for
Gaussian initial conditions \citep[see][for a
review]{bernardeau/etal:2002}, whereas the first term is the effect of
the primordial non-Gaussianity of the ``local type,'' whose
Bardeen's curvature perturbation, $\Phi$, is written as
$\Phi({\mathbi x})=\phi({\mathbi x})+\fnl\phi^2({\mathbi x})$, where
$\phi$ is a Gaussian  field
\citep{salopek/bond:1990,gangui/etal:1994,verde/etal:2000,komatsu/spergel:2001}. The
latest limit on this parameter is $\fnl=38\pm 21$ (68\% CL) \citep{smith/senatore/zaldarriaga:prep}.

However, Sefusatti \& Komatsu's equation, Eq.~(\ref{eq:SK07}), may
require modifications, in light of recent analytical
\citep{dalal/etal:2008,matarrese/verde:2008,slosar/etal:2008,afshordi/tolley:2008,taruya/etal:2008,mcdonald:2008}
and numerical 
\citep{dalal/etal:2008,desjacques/seljak/iliev:prep,pillepich/porciani/hahn:prep,grossi/etal:prep} studies
of the effects of primordial non-Gaussianity on the galaxy power
spectrum. These studies have discovered an unexpected signature of
primordial non-Gaussianity in the form of a {\it scale-dependent
galaxy bias}, i.e., $P_g(k,z)=b_1^2(z)P_m(k,z)\to [b_1(z)+\Delta b(k,z)]^2P_m(k,z)$,
where 
\begin{equation}
 \Delta b(k,z) = \frac{3(b_1(z)-1)\fnl\Omega_mH_0^2\delta_c}{D(z)k^2T(k)},
\label{eq:bk}
\end{equation}
and $\delta_c\simeq1.68$ is the threshold linear density contrast for a
spherical collapse of an overdensity region. 

Then, several questions arise: can we still use Eq.~(\ref{eq:SK07}) for
the bispectrum? Should we replace $b_1$ by $b_1+\Delta b(k)$? Does the
first line in Eq.~(\ref{eq:SK07}) somehow give the same correction as $\Delta
b(k)$? How about $b_2$? We are going to
address these questions in this paper.

\section{Bispectrum of Dark Matter Halos}
In this section, we derive the galaxy bispectrum for non-Gaussian
initial conditions by using two different methods.
In \S\ref{sec:MLB}, we shall use the
``functional integration method'' for computing $n$-point correlation
functions of {\it peaks} of the cosmological density fluctuations
\citep{politzer/wise:1984,grinstein/wise:1986}.
In \S\ref{sec:local_bias}, we shall present an alternative derivation of the 
same result by using a local bias assumption.

\subsection{Mararrese-Lucchin-Bonometto (MLB) method}\label{sec:MLB}

We shall use the Matarrese-Lucchin-Bonometto (MLB) formula
\citep{matarrese/etal:1986} which allows one to calculate the $n$-point
correlation functions of peaks for non-Gaussian 
initial conditions.
This approach is especially well suited for our purposes, as 
 \citet{matarrese/verde:2008} have applied the MLB formula to compute
 the scale-dependent bias of the galaxy power spectrum. We shall apply
 the MLB formula to compute the galaxy bispectrum 
for general non-Gaussian intial condition.

We study the three-point correlation function of the spatial distribution
of dark matter halos.
Let us consider the probability of finding three halos 
within three arbitrary volume elements: $dV_1$, $dV_2$, and $dV_3$,
which are at $\mathbi{x}_1$, $\mathbi{x}_2$, and $\mathbi{x}_3$,
respectively, as \citep{peebles:1980}
\begin{eqnarray}
\nonumber
P(\mathbi{x}_1,\mathbi{x}_2,\mathbi{x}_3)
&=&
\bar{n}^3\left[
1+\xi_h(x_{12})+\xi_h(x_{23})+\xi_h(x_{31})
\right.
\\
&&
\left.
+\zeta_h(\mathbi{x}_1,\mathbi{x}_2,\mathbi{x}_3)
\right]dV_1dV_2dV_3,
\end{eqnarray}
where $x_{ij}\equiv|\mathbi{x}_i-\mathbi{x}_j|$, and
 $\xi_h$ and $\zeta_h$ are the two- and three-point correlation
 functions of halos,  respectively. 

The next step is to relate the correlation functions of halos,
$\xi_h$ and $\zeta_h$, to those of the underlying matter distribution
function. The locations of halos coincide with those of the peaks of the
matter density fluctuations; thus, one can compute  $\xi_h$ and
$\zeta_h$ by computing the correlation functions of peaks above a
certain threshold, above which the peaks collapse into halos.

We shall assume that halos would be formed in the region where the
{\it smoothed} 
linear density contrast exceeds $\delta_c$.  For a spherical collapse in
an Einstein-de Sitter universe $\delta_c=3(12\pi)^{2/3}/20\simeq 1.68$, and one can find other
values in the ellipsoidal collapse  in arbitrary cosmological
models \citep[see, e.g.,][for a review]{cooray/sheth:2002}.
The mass of halos is determined by the smoothing radius, $R$,
i.e., $M=(4\pi/3)\rho_m R^3$, where $\rho_m$ is the average mass density
of the universe. The smoothed density contrast, $\delta_R$, is related
to the underlying mass fluctuations, $\delta_m$, as 
$\delta_R({\mathbi x})=\int d^3{\mathbi x}'W_R(\left|{\mathbi x}-{\mathbi
x}'\right|)\delta_m({\mathbi x}')$, where $W_R(x)$ is a smoothing
function. We shall use a top-hat filter with radius $R$ for $W_R(x)$.

Using the MLB formula, we find
\begin{eqnarray}
&&
\nonumber
1+\xi_h(x_{12})+\xi_h(x_{23})+\xi_h(x_{31})
+\zeta_h(\mathbi{x}_1,\mathbi{x}_2,\mathbi{x}_3)
\\
&=&
\nonumber
\exp
\left[
\frac{1}{2}\frac{\nu^2}{\sigma_R^2}
\sum_{i\neq j}\xi_R^{(2)}(x_{ij})
+
\sum_{n=3}^\infty
\left\{
\sum_{m_1=0}^{n}
\sum_{m_2=0}^{n-m_1}
\frac{\nu^n\sigma_R^{-n}}{m_1!m_2!m_3!}
\right.\right.
\\
&&
\hspace{.8cm}
\nonumber
\times
\xi_R^{(n)}\left(
\begin{array}{ccc}
\mathbi{x}_1,\cdots,\mathbi{x}_1,
&\mathbi{x}_2,\cdots,\mathbi{x}_2,
&\mathbi{x}_3,\cdots,\mathbi{x}_3\\
m_1\,\mathrm{times} & m_2\,\mathrm{times} & m_3\,\mathrm{times}
\end{array}
\right)
\\
\label{eq:P3_nG}
&&
\left.\left.
\hspace{.8cm}
-3\frac{\nu^n\sigma_R^{-n}}{n!}
\xi_R^{(n)}\left(
\begin{array}{c}
\mathbi{x},\cdots,\mathbi{x}\\
n\,\mathrm{times}
\end{array}\right)
\right\}
\right],
\end{eqnarray}
where $m_3\equiv n-m_1-m_2$, $\nu\equiv \delta_c/\sigma_R$,
$\sigma_R^2$ is the variance of matter 
density fluctuations smoothed by a top-hat filter with radius $R$,
and $\xi_R^{(n)}$ denotes the connected parts of the $n$-point 
correlation functions of the underlying matter density fields 
smoothed by a top-hat filter of radius $R$.
Here, we have assumed that we are dealing with high density peaks, i.e.,
$\nu\gg 1$, which are equivalent to highly biased galaxies, $b_1\gg 1$.

As $\xi_R^{(n)}\ll 1$  on the large scales  that we
are interested in, 
we expand the exponential in Eq.~(\ref{eq:P3_nG}). We keep the terms up to the four-point
function, as this term provides the dominant contribution to
the three-point function. We find
\begin{eqnarray}
\nonumber
\zeta_h(\mathbi{x}_1,\mathbi{x}_2,\mathbi{x}_3)
&=&
\frac{\nu^3}{\sigma_R^3}
\xi_R^{(3)}(\mathbi{x}_1,\mathbi{x}_2,\mathbi{x}_3)
\\
\nonumber
&+&
\frac{\nu^4}{\sigma_R^4}
\left[
\xi_R^{(2)}(x_{12})\xi_R^{(2)}(x_{23})
+
(2~\mathrm{cyclic})
\right]
\\
&+&
\frac{\nu^4}{2\sigma_R^4}
\left[
\xi_R^{(4)}(\mathbi{x}_1,\mathbi{x}_1,\mathbi{x}_2,\mathbi{x}_3)
+
(2~\mathrm{cyclic})
\right].
\end{eqnarray}
The bispectrum of halos in Lagrangian space, $B^L_h(\mathbi{k}_1,\mathbi{k}_2,\mathbi{k}_3)$, is the Fourier transform of
$\zeta_h(\mathbi{x}_1,\mathbi{x}_2,\mathbi{x}_3)$:
\begin{eqnarray}
\nonumber
&&B^L_h(\mathbi{k}_1,\mathbi{k}_2,\mathbi{k}_3)
\\
\nonumber
&=&
\frac{\nu^3}{\sigma_R^3}
\biggl[
B_R(\mathbi{k}_1,\mathbi{k}_2,\mathbi{k}_3)
+
\frac{\nu}{\sigma_R}
\left\{P_R(k_1)P_R(k_2)
+(2~\mathrm{cyclic})
\right\}
\\
\label{eq:bisp_nG_L}
&&
+
\frac{\nu}{2\sigma_R}
\int\frac{d^3q}{(2\pi)^3}
T_R(\mathbi{q},\mathbi{k}_1-\mathbi{q},\mathbi{k}_2,\mathbi{k}_3)
+(2~\mathrm{cyclic})\biggl],
\end{eqnarray}
where $T_R$ is the trispectrum, Fourier transform of $\xi_R^{(4)}$.
Here, we call $B_h^L$ the Lagrangian space bispectrum, as it 
 relates the halo over-density to the \textit{initial} matter
 overdensity with its amplitude extrapolated  
 to the present epoch. If we assume that the halos move in the same way
 as matter, the observed bispectrum in Eulerian space, $B_h$, would be the same 
expression with Eq.~(\ref{eq:bisp_nG_L}), except for the coefficients:
\begin{eqnarray}
\nonumber
&&B_h(\mathbi{k}_1,\mathbi{k}_2,\mathbi{k}_3)
\\
\nonumber
&=&
b_1^3
\biggl[
B_R(\mathbi{k}_1,\mathbi{k}_2,\mathbi{k}_3)
+
\frac{b_2}{b_1}
\left\{P_R(k_1)P_R(k_2)
+(2~\mathrm{cyclic})
\right\}
\\
\label{eq:bisp_nG}
&&
+
\frac{\delta_c}{2\sigma_R^2}
\int\frac{d^3q}{(2\pi)^3}
T_R(\mathbi{q},\mathbi{k}_1-\mathbi{q},\mathbi{k}_2,\mathbi{k}_3)
+(2~\mathrm{cyclic})\biggl].
\end{eqnarray}

Here, $b_1$ is the so-called linear Eulerian bias parameter,
$b_1=1+\nu/\sigma_R$,\footnote{Note that this expression,
$b_1=1+\nu/\sigma_R$, agrees with the linear halo bias parameter 
derived by \citet{mo/white:1996}, $b_1=1+(\nu^2-1)/\delta_c$, for high density
peaks, $\nu\gg 1$.} and $b_2$ is the non-linear bias parameter. 

\subsection{Alternative derivation}\label{sec:local_bias}
In this section, we present an alternative derivation of the galaxy 
bispectrum, Eq.~(\ref{eq:bisp_nG}). 
On large enough scales, we may approximate the relation 
between the galaxy distribution and the underlying density 
fluctuation as a local function. We then
Taylor-expand this local function in a power series of $\delta_m$ (see
Eq.~(\ref{eq:def_local_bias})).

When computing the correlation functions of halos of a given mass $M$,
we may smooth 
the matter density field with the same filter over the corresponding
length scale $R$, 
$W_R(|\mathbi{x}-\mathbi{x}'|)$, which was defined in the previous section.
We then Taylor-expand $\delta_g$ in a power series of 
the smoothed density field, $\delta_R(\mathbi{x})$, as
\begin{equation}\label{eq:def_local_bias_R}
 \delta_g({\mathbi x})=b_{1}\delta_R({\mathbi x}) 
+ 
\frac{b_{2}}2
\left[
\delta_R^2({\mathbi x}) - \sigma_R^2
\right]
+\cdots.
\end{equation}
In Fourier space, one finds
\begin{eqnarray}
\nonumber
\delta_g({\mathbi k})
&=&b_1\delta_R({\mathbi k}) 
\\
\label{eq:Fourier_local_bias_R}
&&
\nonumber
+ \frac{b_2}2
\biggl[
\int\frac{d^3q}{(2\pi)^3}
\delta_R({\mathbi k}-{\mathbi q})\delta_R({\mathbi q})
-\sigma_R^2\delta^D(\mathbi{k})
\biggl]\\
&&
+\cdots,
\end{eqnarray}
where $\delta^D$ is the Dirac delta function.
We calculate the bispectrum of galaxies 
directly from Eq.~(\ref{eq:Fourier_local_bias_R}):
\begin{eqnarray}
\nonumber
&&
\langle
\delta_g(\mathbi{k}_1)
\delta_g(\mathbi{k}_2)
\delta_g(\mathbi{k}_3)
\rangle
\\
\nonumber
&=&
b_1^3
\langle
\delta_R(\mathbi{k}_1)
\delta_R(\mathbi{k}_2)
\delta_R(\mathbi{k}_3)
\rangle
\\
\nonumber
&&+
\frac{b_1^2b_2}{2}
\biggl[
\int\frac{d^3q}{(2\pi)^3}
\langle
\delta_R(\mathbi{k}_1-\mathbi{q})
\delta_R(\mathbi{q})
\delta_R(\mathbi{k}_2)
\delta_R(\mathbi{k}_3)
\rangle
\\
\label{eq:bisp_1st}
&&-\sigma_R^2\delta^D(\mathbi{k}_1)
\langle
\delta_R(\mathbi{k}_2)
\delta_R(\mathbi{k}_3)
\rangle
+(2~\mathrm{cyclic})
\biggl].
\end{eqnarray}
The first term of Eq.~(\ref{eq:bisp_1st}) is the matter bispectrum,
$$
\langle
\delta_R(\mathbi{k}_1)
\delta_R(\mathbi{k}_2)
\delta_R(\mathbi{k}_3)
\rangle
=(2\pi)^3B_R(k_1,k_2,k_3)\delta^D(\mathbi{k}_{123}),
$$
where $\mathbi{k}_{123}\equiv\mathbi{k}_1+\mathbi{k}_2+\mathbi{k}_3$.
We further calculate the ensemble average of the four-point function 
in the second term of Eq.~(\ref{eq:bisp_1st}). 
For non-Gaussian density fields, four-point function is given by
a sum of connected (trispectrum) and unconnected 
(products of the power spectra) parts as
\begin{eqnarray*}
&&\langle
\delta_R(\mathbi{k}_1-\mathbi{q})
\delta_R(\mathbi{q})
\delta_R(\mathbi{k}_2)
\delta_R(\mathbi{k}_3)
\rangle
\\
&=&
(2\pi)^6 
P_R(q)P_R(k_2)
\delta^D(\mathbi{k}_1) 
\delta^D(\mathbi{k}_2+\mathbi{k}_3)
\\
&&+
(2\pi)^6 
P_R(k_2)P_R(k_3)
\delta^D(\mathbi{k}_2+\mathbi{q}) 
\delta^D(\mathbi{k}_3+\mathbi{k}_1-\mathbi{q})
\\
&&+
(2\pi)^6 
P_R(k_2)P_R(k_3)
\delta^D(\mathbi{k}_3+\mathbi{q}) 
\delta^D(\mathbi{k}_2+\mathbi{k}_1-\mathbi{q})
\\
&&+
(2\pi)^3 
T_R(\mathbi{k}_1-\mathbi{q},\mathbi{q},\mathbi{k}_2,\mathbi{k}_3)
\delta^D(\mathbi{k}_{123}),
\end{eqnarray*}
where $T_R$ is the matter trispectrum.
Note that the first term in the above equation cancels the last
term in Eq.~(\ref{eq:bisp_1st}). Combining the above equations,
Eq.~(\ref{eq:bisp_1st}) becomes
\begin{eqnarray}
\nonumber
&&
B_h(\mathbi{k}_1,\mathbi{k}_2,\mathbi{k}_3)
\\
\nonumber
&=&
b_1^3
\biggl[
B_R(\mathbi{k}_1,\mathbi{k}_2,\mathbi{k}_3)
+\frac{b_2}{b_1}\left\{P_R(k_1)P_R(k_2)+(2~\mathrm{cyclic})\right\}
\\
\label{eq:bisp_nG1}
&&+
\frac{1}{2}\frac{b_2}{b_1}
\int\frac{d^3q}{(2\pi)^3}
T_R(\mathbi{q},\mathbi{k}_1-\mathbi{q},\mathbi{k}_2,\mathbi{k}_3)
+(2~\mathrm{cyclic})
\biggl].
\end{eqnarray}

Therefore, we find that the MLB method and the locality bias assumption
give formally the same results. This may imply that the process of picking up 
density peaks over a certain threshold \textit{is} the local process.

At the same time, there is a subtle difference between 
Eq.~(\ref{eq:bisp_nG}) and Eq.~(\ref{eq:bisp_nG1}): the coefficient in the
last term. By evaluating the last cyclic term in Eq.~(\ref{eq:bisp_nG1})
for the local type of non-Gaussianity, we find that 
the integration of the smoothed trispectrum depends
on the smoothing scale, $R$, up to a constant factor of $1/\sigma_R^2$ on 
large scales, say, $k<0.1~h/\mathrm{Mpc}$. 
For example, the bottom right panel of Fig.~\ref{fig:z0f40} shows 
that $B_{\fnl^2}^{nG}$ and $B_{\gnl}^{nG}$, which are defined in 
Eq.~(\ref{eq:bfnl2*}) and Eq.~(\ref{eq:bgnl}), respectively, 
do not depend on the smoothing scale $R$, 
as they include $1/\sigma_R^2$ in their definitions.

Therefore, it is physically more sensible to include $\sigma_R^2$
explicitly in the equation such that the  dependence  on smoothing scales
 on large scales can be absorbed by the bias parameters.
This motivates our writing the final form of the halo bispectrum,
derived from the local bias assumption, as
\begin{eqnarray}
\nonumber
&&
B_h(\mathbi{k}_1,\mathbi{k}_2,\mathbi{k}_3)
\\
\nonumber
&=&
b_1^3
\biggl[
B_R(\mathbi{k}_1,\mathbi{k}_2,\mathbi{k}_3)
+\frac{b_2}{b_1}\left\{P_R(k_1)P_R(k_2)+(2~\mathrm{cyclic})\right\}
\\
\label{eq:bisp_nG2}
&&
\nonumber
+
\frac{\tilde{b}_2}{b_1}
\frac{1}{2\sigma_R^2}
\int\frac{d^3q}{(2\pi)^3}
T_R(\mathbi{q},\mathbi{k}_1-\mathbi{q},\mathbi{k}_2,\mathbi{k}_3)
+(2~\mathrm{cyclic})
\biggl],\\
\end{eqnarray}
with three bias parameters: $b_1$, $b_2$ and $\tilde{b}_2\equiv b_2\sigma_R^2$.
Note that $\tilde{b}_2/b_1\to \delta_c$ for 
the MLB formula. Although $\delta_c$ is known to be 
$1.68$ for the spherically collapsed halo in the flat matter dominated 
universe, its precise value, in this context, needs to be tested 
against N-body simulations.

Eqs.~(\ref{eq:bisp_nG}) and (\ref{eq:bisp_nG2}) are
the first main results of this paper, which are general and can be
applied to  any models of non-Gaussianities, once we know the bispectrum
and trispectrum of the underlying matter density field.
Note that Eq.~(\ref{eq:bisp_nG2}) was also obtained independently 
by \citet{sefusatti:prep}.

In principle both $b_1$ and $b_2$
are calculable from the theory of collapse of halos 
\citep[see, e.g.,][for a review]{cooray/sheth:2002};
however, in practice it is more convenient and safe to treat them as
free parameters that one should marginalize over when extracting the
cosmological information, such as $\fnl$. See
\citet{jeong/komatsu:2009} for the same argument applied to the galaxy
power spectrum.

\section{Effects of Local-type Primordial non-Gaussianity on The Halo Bispectrum}

In this section we shall evaluate Eq.~(\ref{eq:bisp_nG2}) for the
local-type primordial non-Gaussianity with a high-order term added: 
\begin{equation}
\Phi(\mathbi{x})
=
\phi(\mathbi{x}) 
+ \fnl\left[\phi^2(\mathbi{x})-\langle\phi^2\rangle\right]
+ \gnl\phi^3(\mathbi{x}).
\label{eq:localNG}
\end{equation}
The cubic-order term does not generate the bispectrum of CMB anisotropy
or the matter density fluctuations at the leading order; however, it
does generate the trispectrum, and the CMB trispectrum has been calculated
by \citet{okamoto/hu:2002,kogo/komatsu:2006}.
On the other hand, the bispectrum of {\it halos} receives the
contribution from the trispectrum (see the last term in
Eq.~(\ref{eq:bisp_nG2})), and thus it is necessary to include
$\gnl$. 

To calculate various components of the bispectra shown in
Eq.~(\ref{eq:bisp_nG2}), we calculate
the transfer function, $T(k)$, and the power spectra with the
cosmological parameters in Table 1 (``WMAP+BAO+SN'') of
\citet{komatsu/etal:2009}. 

As for the smoothing scale, we use $R=1~h^{-1}~\mathrm{Mpc}$.
Although the smoothing scale explicitly appears in the equation, 
it makes negligible differences for the bispectrum on large scales,
$k\ll 1/R$.

Note that we shall adopt the non-standard convention in which 
$\Phi(\mathbi{x})$ is  Bardeen's curvature perturbation
extrapolated to the present epoch, $z=0$, using the linear growth factor of
$\Phi$, $g(z)\equiv (1+z)D(z)$. Therefore, our $\fnl$ and $\gnl$ in this
paper are different from those in the literature on the CMB
non-Gaussianity by a factor of $g(1090)/g(0)$, i.e., 
$\fnl=[g(1090)/g(0)]\fnl^{\rm CMB}$ and 
$\gnl=[g^2(1090)/g^2(0)]\gnl^{\rm CMB}$
\footnote{
The ratio $g(1090)/g(0)$ is $1.308$ for the cosmological parameters 
in Table 1 (``WMAP+BAO+SN'') of \citet{komatsu/etal:2009}. 
}.

The bispectrum and trispectrum of $\Phi$ are given by
\begin{equation}
B_\Phi(\mathbi{k}_1,\mathbi{k}_2,\mathbi{k}_3)
=2\fnl
\left[
P_\phi(k_1)P_\phi(k_2) + (2~\mathrm{cyclic})
\right],
\label{eq:bphi}
\end{equation}
and
\begin{eqnarray}
\nonumber
&&T_\Phi(\mathbi{k}_1,\mathbi{k}_2,\mathbi{k}_3,\mathbi{k}_4)
\\
\nonumber
&
=
&
6\gnl 
\left[
P_\phi(k_1) P_\phi(k_2) P_\phi(k_3) + (3~\mathrm{cyclic})
\right]
+2\fnl^2
\\
&&\times
\left[
P_\phi(k_1) P_\phi(k_2)
\left\{
 P_\phi(k_{13})
+ P_\phi(k_{14})
\right\}
+
(11~\mathrm{cyclic})
\right],
\label{eq:tphi}
\end{eqnarray}
with $k_{ij}=|\mathbi{k}_i+\mathbi{k}_j|$, respectively.

While Eq.~(\ref{eq:tphi}) is the consequence of Eq.~(\ref{eq:localNG}),
general multi-field inflation models do not necessarily relate the
coefficients of the trispectrum to that of the bispectrum. Therefore,
one may generalize Eq.~(\ref{eq:tphi}) by replacing $\fnl^2$ with a new
parameter, $\tau_{\rm NL}$, which may or may not be related to $\fnl$:
\begin{eqnarray}
\nonumber
&&T_\Phi(\mathbi{k}_1,\mathbi{k}_2,\mathbi{k}_3,\mathbi{k}_4)
\\
\nonumber
&
=
&
6\gnl 
\left[
P_\phi(k_1) P_\phi(k_2) P_\phi(k_3) + (3~\mathrm{cyclic})
\right]
+\frac{25}{18}\tau_{\rm NL}
\\
&&\times
\left[
P_\phi(k_1) P_\phi(k_2)
\left\{
 P_\phi(k_{13})
+ P_\phi(k_{14})
\right\}
+
(11~\mathrm{cyclic})
\right].
\label{eq:tphi2}
\end{eqnarray}
Note that the coefficient of $\tau_{\rm NL}$ reflects the definition of
$\tau_{\rm NL}$ introduced by \citet{boubekeur/lyth:2006}.
This opens up an exciting possibility that the galaxy bispectrum can
test whether $\tau_{\rm NL}=(6\fnl/5)^2$ or other predictions for the
relation between $\tau_{\rm NL}$ and $\fnl$ are satisfied observationally.

We transform these spectra to those of the smoothed linear density contrasts,
using the  Poisson equation,
\begin{equation}
\delta^{(1)}_R(k) 
=
\frac{2}{3}\frac{k^2T(k)}{H_0^2\Omega_{m}}\tilde{W}_R(k)\Phi(k)
\equiv
\Mr(k)\Phi(k),
\end{equation}
where $\tilde{W}_R(k)$ is the Fourier transform
of the top-hat filter with radius $R$. Note that
$\tilde{W}_R(k)\to 1$ as $k\to 0$. In general $\tilde{W}_R(k)\simeq 1$
for $k\ll 1/R$.
Then, we calculate the $n$-point function of the 
matter density fields from the corresponding correlator of curvature 
perturbations by
$$
\langle
\delta_R^{(1)}(\mathbi{k}_1)
\cdots
\delta_R^{(1)}(\mathbi{k}_n)
\rangle
=
\prod_{i=1}^n
\Mr(k_i)
\langle
\Phi(\mathbi{k}_1)
\cdots
\Phi(\mathbi{k}_n)
\rangle.
$$

\subsection{Known Terms}
\subsubsection{Formula}
The first term in Eq.~(\ref{eq:bisp_nG2}) 
contains the bispectrum of matter density fluctuations, $B_R$,
which consists of  two pieces:
(i) the non-linear evolution of gravitational clustering
($B_{m}^G$)
and (ii) primordial non-Gaussianity ($B_{\fnl}^{nG0}$)\footnote{
We ignore the following term in $B_R(k_1,k_2,k_3)$:
\begin{align*}
\prod_{i=1}^3 \tilde{W}_R(k_i)
\int\frac{d^3q}{(2\pi)^3}
F_2^{(s)}(\mathbi{q},\mathbi{k}_1-\mathbi{q})
T(\mathbi{q},\mathbi{k}_1-\mathbi{q},\mathbi{k}_2,\mathbi{k}_3)
+(2~\mathrm{cyclic}),
\end{align*}
where $T$ is the unfiltered primordial trispectrum which contains
$\fnl^2$ and $\gnl$. This term is negligibly small \citep{scoccimarro/etal:2004}.
}:
\begin{equation}
B_R(k_1,k_2,k_3)
=
B_{m}^G(k_1,k_2,k_3)
+
\fnl B_{\fnl}^{nG0}(k_1,k_2,k_3),
\end{equation}
where
\begin{eqnarray}
\nonumber
B_{m}^G(k_1,k_2,k_3)
&\equiv& 
\tilde{W}_R(k_1)\tilde{W}_R(k_2)\tilde{W}_R(k_3)
2F_2^{(s)}(\mathbi{k}_1,\mathbi{k}_2)
\\
&&\times
P_m(k_1)P_m(k_2)
+(2~\mathrm{cyclic}),
\label{eq:bmg}
\end{eqnarray}
with $F_2^{(s)}$ given by Eq.~(\ref{eq:F2}), and 
\begin{eqnarray}
\nonumber
& &B_{\fnl}^{nG0}(k_1,k_2,k_3)\\
\nonumber
&\equiv& 2\prod_{i=1}^3 \Mr(k_i)
\left[
P_\phi(k_1)P_\phi(k_2)+ (2~\mathrm{cyclic})
\right]\\
&=&
2\frac{P_R(k_1)}{\Mr(k_1)}\frac{P_R(k_2)}{\Mr(k_2)}\Mr(k_3)+
(2~\mathrm{cyclic}).
\label{eq:bfnl}
\end{eqnarray}
One finds that Eqs.~(\ref{eq:bfnl}) and (\ref{eq:bmg}) agree with the
first and the second terms in Eq.~(\ref{eq:SK07}) on the scales much
larger than the smoothing scale, i.e., $k\ll 1/R$, for which
$\tilde{W}_R\to 1$.

One might wonder if it is OK to include
the bispectrum from non-linear evolution of density fluctuations in the
MLB formula, as Eq.~(\ref{eq:bisp_nG}) is usually used for the Lagrangian density
fluctuations, i.e., ``initial'' fluctuations. However, it is perfectly
OK to use the evolved density fluctuations in this formula, as one can
always use the evolved density fluctuations as the initial data. For
example, we can think of starting our calculation at $z=10$, and ask the
MLB formula to take the initial condition at $z=10$, including non-linear
correction. Since we know how to compute the bispectrum,
trispectrum, etc., of the underlying mass distribution at $z=10$
(including non-linear effects), we can use this information in the MLB formula.
In other words, the ``initial'' distribution does not need to be
primordial. We can provide the evolved density field as the initial
data, and compute the peak statistics. The MLB formula 
does not care whether the source of non-Gaussianity
is truly primordial or not: the only conditions that we must respect for
Eq.~(\ref{eq:bisp_nG}) to be valid are (i) high peaks ($\nu\gg 1$), and
(ii) $n$-point correlation functions are much less than unity,
$\xi_R^{(n)}\ll 1$, so that the exponential in Eq.~(\ref{eq:P3_nG}) can be
Taylor-expanded. In this case one would lose an ability to calculate the
bias parameters, $b_1$ and $b_2$, using, e.g., a halo model; however,
this is not a disadvantage, as the halo model calculations of the {\it
galaxy} bias parameter, $b_1$ and
$b_2$, are at best qualitative even for Gaussian initial conditions 
\citep[see, e.g.,][]{jeong/komatsu:2009}. In our approach the
coefficients of individual terms in 
Eqs.~(\ref{eq:bisp_nG}) and (\ref{eq:bisp_nG2}), 
including $\delta_c$, are free parameters, and need to
be determined from observations themselves.

\subsubsection{Shape Dependence: 
Non-linear Gravitational Evolution and Non-linear Galaxy Bias}

How about the shape dependence? 
First, let us review the structure of $B_m^G(k_1,k_2,k_3)$
(Eq.~(\ref{eq:bmg})), which has been studied extensively in the
literature \citep[see][for a review]{bernardeau/etal:2002}.

\begin{figure*}
\begin{center}
\includegraphics[width=16cm]{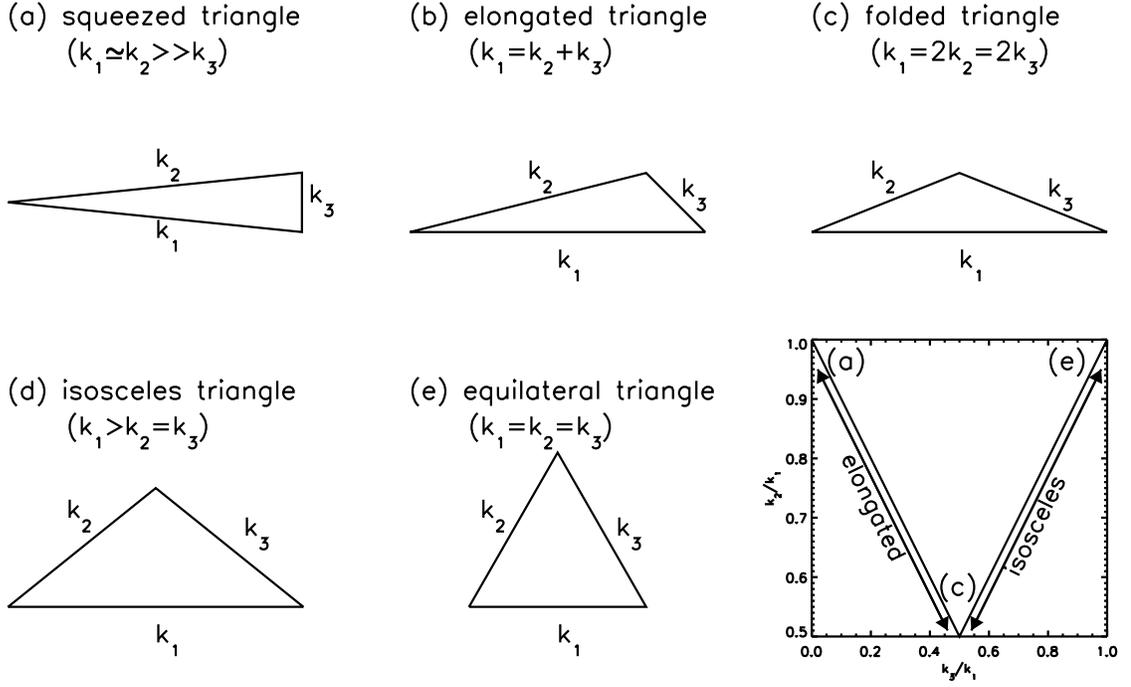}
\caption{
 Visual representations of triangles forming the bispectrum,
 $B(k_1,k_2,k_3)$, with various combinations of
 wavenumbers satisfying $k_3\le k_2\le k_1$.
}
\label{fig:triangles}
\end{center}
\end{figure*}

Here, let us offer a useful way of visualizing the shape
dependence of the bispectrum.  
We can study the structure of the bispectrum 
by plotting the magnitude of $B_m^G$ as a
function of $k_2/k_1$ and $k_3/k_1$ for a given $k_1$, with 
a condition that $k_1\ge k_2\ge k_3$ is satisfied. 
In order to classify various shapes of the triangles, 
let us use the following names:
squeezed ($k_1\simeq k_2\gg k_3$), elongated ($k_1=k_2+k_3$), 
folded ($k_1=2k_2=2k_3$), isosceles ($k_2=k_3$), and equilateral
($k_1=k_2=k_3$). See (a)--(e) of Fig.~\ref{fig:triangles} for the visual
representations of these triangles. 

\begin{figure*}
\begin{center}
\includegraphics[width=18cm]{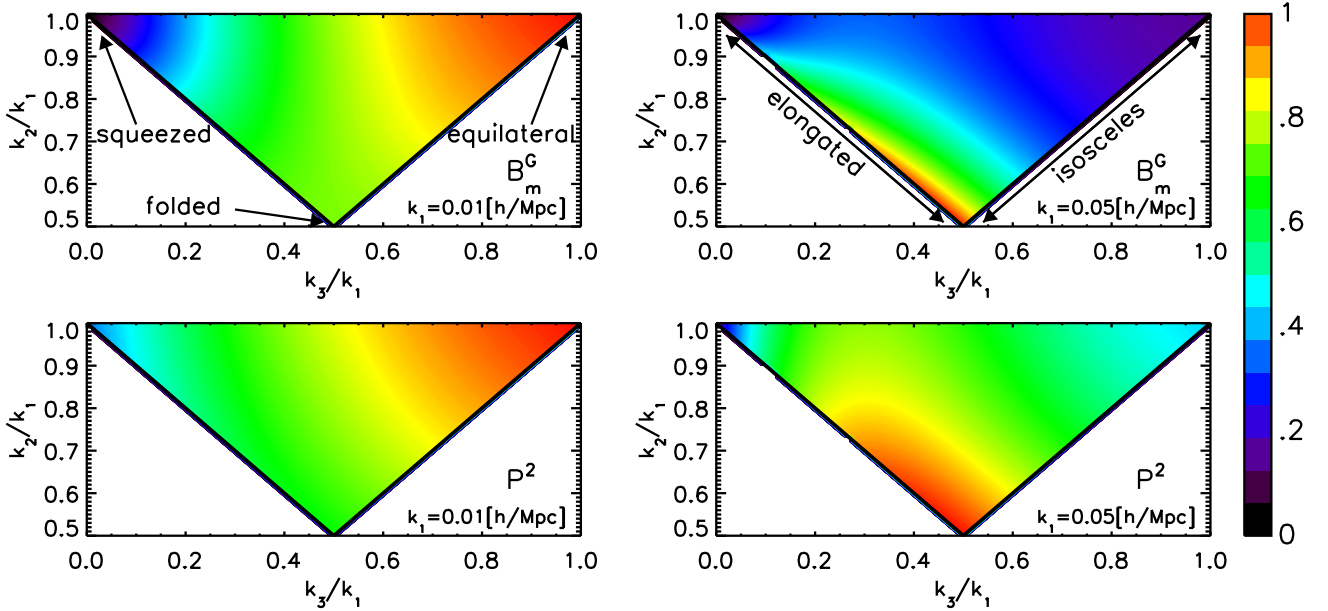}
\caption{
 Shape of the bispectrum, $B(k_1,k_2,k_3)$. Each panel shows the
 amplitude of the bispectrum as a function of $k_2/k_1$ and $k_3/k_1$
 for a given $k_1$, with a condition that $k_3\le k_2\le k_1$ is satisfied.
 The amplitude is normalized
 such that it is unity at the point where the bispectrum takes on the
 maximum value. For the visual representations of the triangle names
 such as the squeezed, elongated, folded, isosceles, and equilateral,
 see Fig.~\ref{fig:triangles}.
(Top Left) The bispectrum from the non-linear
 gravitational evolution, $B_m^G$ (Eq.~(\ref{eq:bmg})), for
 $k_1=0.01~h~{\rm Mpc}^{-1}$. (Top Right)
 $B_m^G$ for  $k_1=0.05~h~{\rm Mpc}^{-1}$.
 (Bottom Left) The bispectrum from the non-linear galaxy biasing,
 $P_R(k_1)P_R(k_2)+(2~\mbox{cyclic})$ (the second term in
 Eq.~(\ref{eq:bisp_nG2})), for $k_1=0.01~h~{\rm Mpc}^{-1}$.
 (Bottom Right) $P_R(k_1)P_R(k_2)+(2~\mbox{cyclic})$ for $k_1=0.05~h~{\rm
 Mpc}^{-1}$. 
         }%
\label{fig:bkG}
\end{center}
\end{figure*}

The top-left panel of Fig.~\ref{fig:bkG} shows
$B_m^G$ for $k_1=0.01~h~{\rm Mpc}^{-1}$. In this regime 
$P_R(k_1)$ is an increasing function of $k_1$ (recall that $P_R(k)$ peaks
at $k\approx 0.02~h~{\rm Mpc}^{-1}$). Let us then pick the first term in
Eq.~(\ref{eq:bmg}),
$F_2^{(s)}({\mathbi k}_1,{\mathbi k}_2)P_R(k_1)P_R(k_2)$, and ignore the
cyclic terms for the moment. 
It follows from Eq.~(\ref{eq:F2}) and the descriptions below it that
$F_2^{(s)}({\mathbi k}_1,{\mathbi k}_2)$
vanishes in the squeezed limit ($\mathbi{k}_1=-\mathbi{k}_2$)
and reaches the maximum in the opposite limit
($\mathbi{k}_1=\alpha\mathbi{k}_2$). Therefore, we would  
expect this term to give large
signals in the ``elongated configurations,'' $k_1=k_2+k_3$; however,
as $P_R(k)$ at $k\lesssim 0.02~h~{\rm Mpc}^{-1}$ is
an increasing function of $k$, one can also get large signals when
$k_1$ and $k_2$ are equally large, $k_1=k_2$. As we have zero signal in
the squeezed limit, $k_3=0$, it follows that we can find a large signal
in the equilateral configuration, $k_1=k_2=k_3$.
A similar argument also applies to the  cyclic terms.
As a result, for $k_1=0.01~h~{\rm Mpc}^{-1}$, we find the largest signal
in the equilateral configuration, and then the signal decreases as we
approach the squeezed configuration, i.e., the signal decreases as we go
from (e) to (a) in Fig.~\ref{fig:triangles}.

The top-right panel of Fig.~\ref{fig:bkG} shows
$B_m^G$ for $k_1=0.05~h~{\rm Mpc}^{-1}$. In this regime 
$P_R(k_1)$ is a {\it decreasing} function of $k_1$, and thus 
the equilateral configurations are no longer as important as 
the folded ones. Instead we have large signals in the folded 
configurations as well as in the elongated  configurations. 
Note that the exact squeezed limit is
 still suppressed due to the form of $F_2^{(s)}$.

In summary, $B_m^G$ usually has the largest signal in the folded and elongated
(or equilateral, depending on the wavenumber)
configurations,
with the squeezed configurations suppressed relative to the others.
The suppression of the squeezed configurations is a generic signature of
the causal mechanism such as the non-linear gravitational evolution that 
$F_2^{(s)}$ describes.

The bispectrum from the non-linear bias term, the second term in
Eq.~(\ref{eq:bisp_nG}), has the same structure as $B_m^G$, but it does
not contain $F_2^{(s)}$. As a result the non-linear bias term does not
have as much suppression  as $B_m^G$
has in the squeezed configuration. 
In addition, as $F_2^{(s)}$ enhances the signal in the elongated
configurations, the non-linear bias term does not
have as much enhancement  as $B_m^G$
has in the elongated configurations. These properties explain the bottom
panels of Fig.~\ref{fig:bkG}.

As $B_m^G$ and the non-linear bias term contain products of
$P_R(k_1)P_R(k_2)$ and the 
cyclic terms, it is often more convenient to deal with $Q_h(k_1,k_2,k_3)$
given by \citep{peebles:1980}
\begin{equation}
 Q_h(k_1,k_2,k_3)\equiv 
\frac{B_h(k_1,k_2,k_3)}{P_R(k_1)P_R(k_2)+(2~\mbox{cyclic})},
\label{eq:qh}
\end{equation}
to reduce the  dependence on the shape of the power spectrum.
This combination is constant and equal to $b_1^2b_2$ for the non-linear
bias term (see the second term in
Eq.~(\ref{eq:bisp_nG})). 

\begin{figure*}
\begin{center}
\includegraphics[width=18cm]{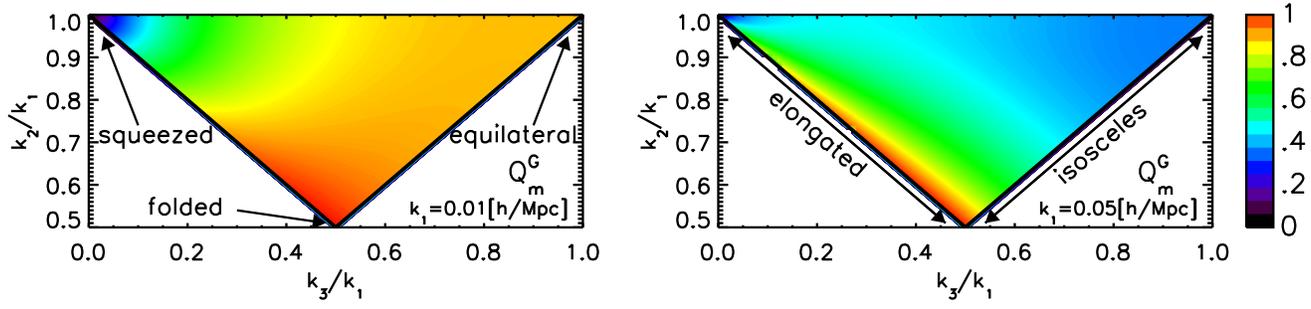}
\caption{
 Same as the top panels of Fig.~\ref{fig:bkG}, but for
 $B_m^G/[P_R(k_1)P_R(k_2)+(2~\mbox{cyclic})]$ (Eq.~(\ref{eq:qh})).
         }%
\label{fig:qkG}
\end{center}
\end{figure*}

The left and right panels of Fig.~\ref{fig:qkG} show
$B_m^G(k_1,k_2,k_3)/[P_R(k_1)P_R(k_2)+(2~\mbox{cyclic})]$ for
$k_1=0.01~h~{\rm Mpc}^{-1}$ and $0.05~h~{\rm Mpc}^{-1}$, respectively.
We find that $Q_h$ better reflects the shape dependence of $F_2^{(s)}$
irrespective of $k_1$: it has the largest signal in the folded and elongated
configurations in both large and small scales. The squeezed
configurations are still heavily suppressed relative to the others.

\begin{figure*}
\begin{center}
\includegraphics[width=18cm]{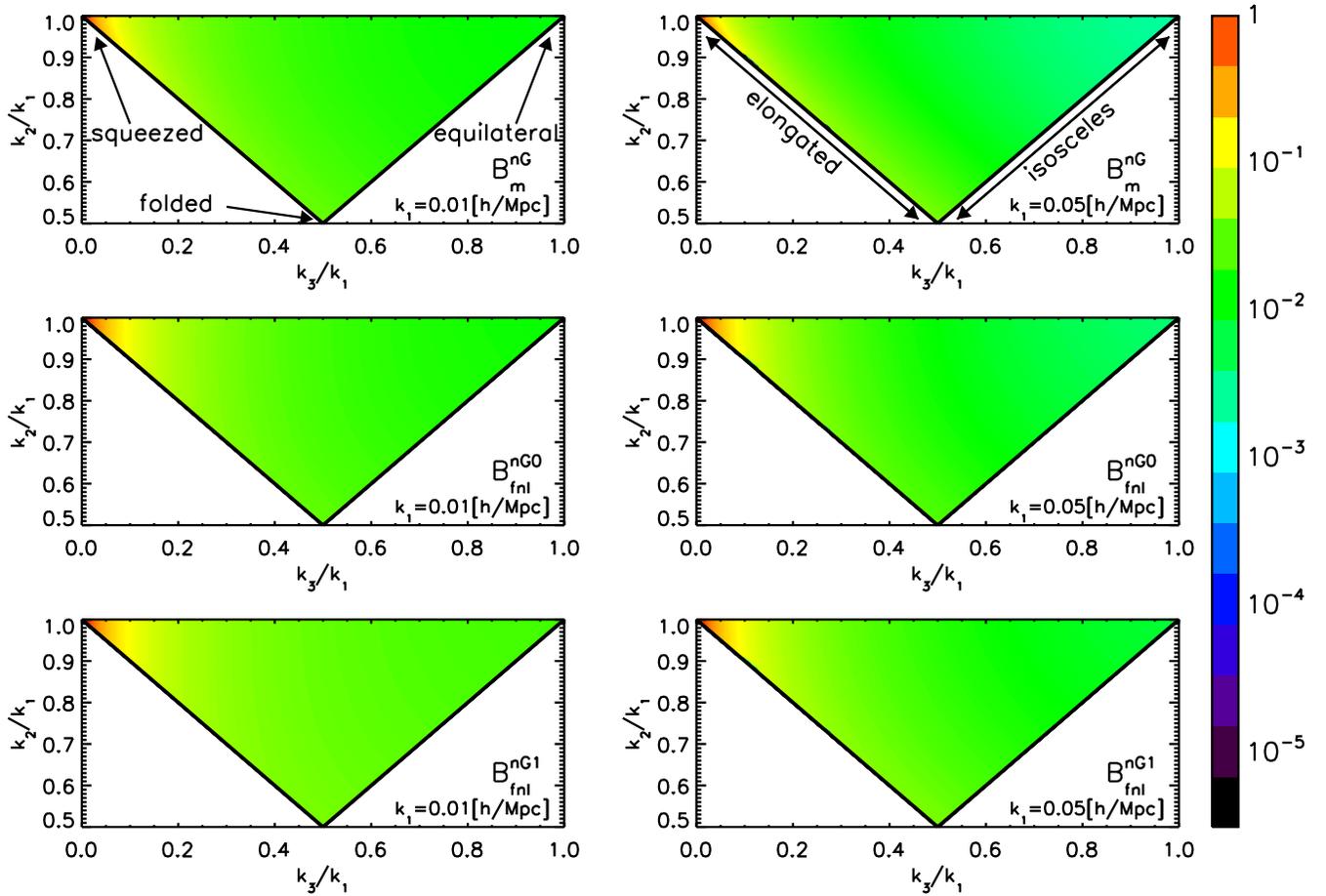}
\caption{
 Same as Fig.~\ref{fig:bkG}, but for
 the terms proportional to $\fnl$.
 (Top) the $B_{m}^{nG}$ term (Eq.~(\ref{eq:BmnG})), 
 (Middle) the $B_{\fnl}^{nG0}$ term (Eq.~(\ref{eq:bfnl})), and 
 (Bottom) the $B_{\fnl}^{nG1}$ term (Eq.~(\ref{eq:BfnlnG1})).
 Note that the non-Gaussian terms diverge in the exact squeezed limit,
 $k_3\to 0$; thus, we show these terms normalized to be unity 
 at  $k_3/k_1=10^{-2}$.
 In order to facilitate the comparison better, we draw the dotted contour
 for all six panels.
         }%
\label{fig:bkNG_fnl}
\end{center}
\end{figure*}
\begin{figure*}
\begin{center}
\includegraphics[width=18cm]{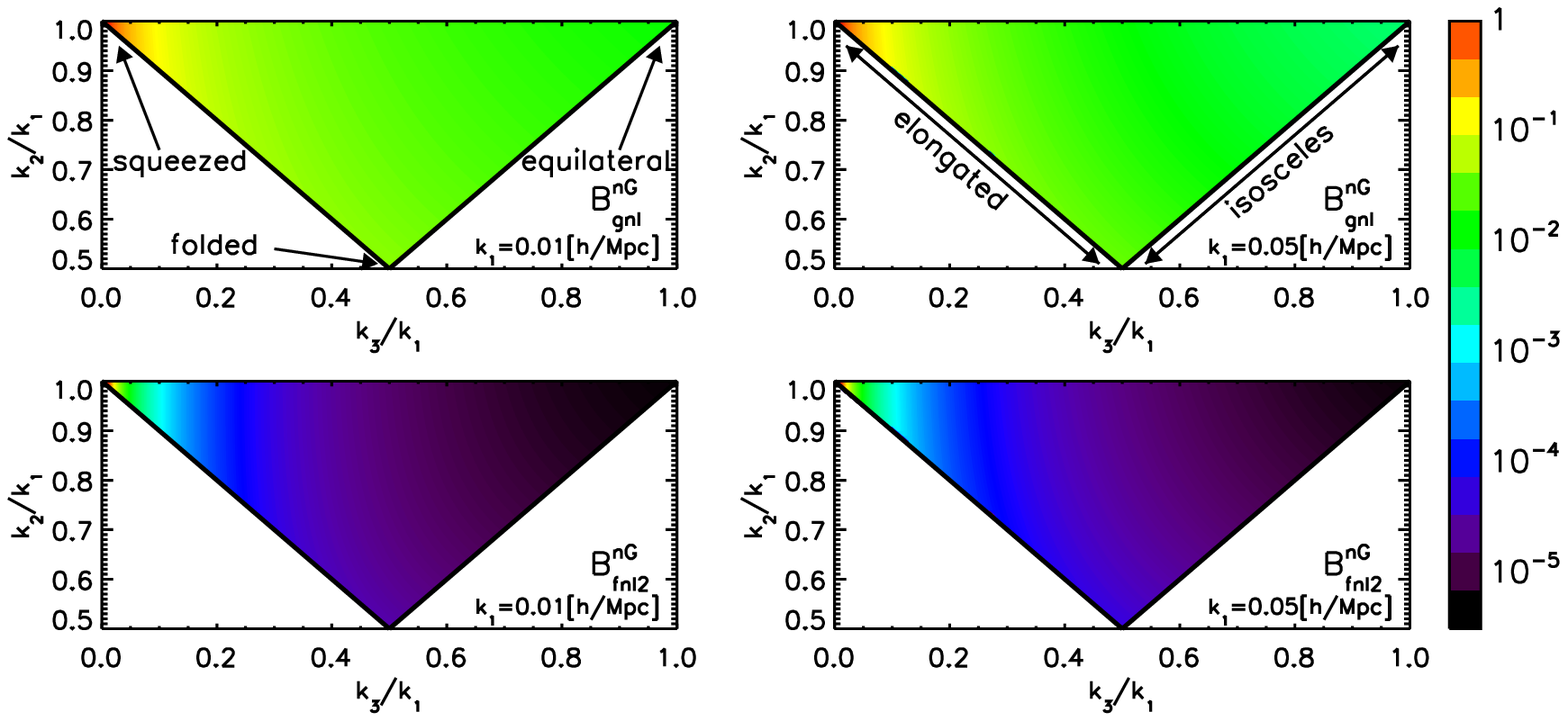}
\caption{
 Same as Fig.~\ref{fig:bkG}, but for
 (Top) the $\gnl$ term (Eq.~(\ref{eq:bgnl})), and 
 (Bottom) the $\fnl^2$ term (Eq.~(\ref{eq:bfnl2})).
 Note that the non-Gaussian terms diverge in the exact squeezed limit,
 $k_3\to 0$; thus, we show these terms normalized to be unity 
 at  $k_3/k_1=10^{-2}$.
 In order to facilitate the comparison better, we draw the dotted contour
 for top panels.
         }%
\label{fig:bkNG}
\end{center}
\end{figure*}

\subsubsection{Shape Dependence: $\fnl$ Term}

How about the $\fnl$ term, $B_{\fnl}^{nG0}(k_1,k_2,k_3)$? 
This term has a completely different structure.
Let us pick the first term,
$\Mr(k_1)P_\phi(k_1)\Mr(k_2)P_\phi(k_2)\Mr(k_3)$, in Eq.~(\ref{eq:bfnl}).
The important point is that the power spectrum of $\phi$ is always a
decreasing function of $k$, i.e., $P_\phi(k)\propto 1/k^3$ for a
scale-invariant spectrum. On the other hand, on large scales we have
$T(k)\to 1$ and $\Mr(k)\propto k^2$. Therefore, collecting 
all the cyclic terms, we find 
$B_{\fnl}^{nG0}(k_1,k_2,k_3)\propto 
k_3^2/(k_1k_2)+k_2^2/(k_1k_3)+k_1^2/(k_2k_3)
=(k_1^3+k_2^3+k_3^3)/(k_1k_2k_3)$. In other words, it has the largest
signal when one of $k$'s is very small, i.e., the squeezed
configurations, which is opposite to the structures of $B_m^G$
and the non-linear bias term.
The middle panels of Fig.~\ref{fig:bkNG_fnl} show $B_{\fnl}^{nG0}$
for $k_1=0.01~h~{\rm Mpc}^{-1}$ and $0.05~h~{\rm Mpc}^{-1}$, and 
we find the largest signals in the squeezed configurations.
We also find that $Q_h$ from the $\fnl$ term, 
$B_{\fnl}^{nG0}(k_1,k_2,k_3)/[P_R(k_1)P_R(k_2)+(2~\mbox{cyclic})]$, still has 
the largest signal in the squeezed configurations. 

These properties allow us to distinguish between the primordial
non-Gaussianity and the other effects such as the non-linear
gravitational evolution and non-linear bias. 
\citet{sefusatti/komatsu:2007} have studied in detail how well one
can separate these effects using $Q_h$.

\subsection{New Term}

\subsubsection{Formula}
Now, we shall evaluate the new term, the third term in
Eq.~(\ref{eq:bisp_nG2}), which was not considered in
\citet{sefusatti/komatsu:2007}. The trispectrum is generated by the
primordial non-Gaussianity, as well as by the non-linear evolution of
matter density fluctuations.
The non-linear evolution of matter density fluctuations 
on large scales is given by perturbation theory 
\citep[see][for a review]{bernardeau/etal:2002}.
Let us expand the filtered non-linear matter 
density field in Fourier space as
\begin{equation}
\delta_R(\mathbi{k})
=
\tilde{W}_R(k)
\left[
\delta^{(1)}(\mathbi{k})
+
\delta^{(2)}(\mathbi{k})
+
\delta^{(3)}(\mathbi{k})
+\cdots
\right],
\end{equation}
where $\delta^{(n)}(\mathbi{k})$ is the $n$-th order quantity of the linear
density contrast, $\delta^{(1)}(\mathbi{k})$.
Then, the connected matter density trispectrum is given by
\begin{eqnarray}
\nonumber
&&T_R(\mathbi{k}_1,\mathbi{k}_2,\mathbi{k}_3,\mathbi{k}_4)
\\
\nonumber
&=&
T_R^{1111}(\mathbi{k}_1,\mathbi{k}_2,\mathbi{k}_3,\mathbi{k}_4)
+\left\{
T_R^{1112}(\mathbi{k}_1,\mathbi{k}_2,\mathbi{k}_3,\mathbi{k}_4)
+(3~\mathrm{cyclic})
\right\}
\\
\nonumber
&&
+
\left\{
T_R^{1113}(\mathbi{k}_1,\mathbi{k}_2,\mathbi{k}_3,\mathbi{k}_4)
+(3~\mathrm{cyclic})
\right\}
\\
\label{eq:def_TR}
&&+
\left\{
T_R^{1122}(\mathbi{k}_1,\mathbi{k}_2,\mathbi{k}_3,\mathbi{k}_4)
+(5~\mathrm{cyclic})
\right\}
+\mathcal{O}(\phi^8),
\end{eqnarray}
with $T_R^{ijkl}$ given by
\begin{eqnarray}
\nonumber
&&
(2\pi)^3\delta^D(\mathbi{k}_{1234})
T_R^{ijkl}(\mathbi{k}_1,\mathbi{k}_2,\mathbi{k}_3,\mathbi{k}_4)
\\
&\equiv&
\prod_{n=1}^4
\tilde{W}_R(k_n)
\langle
\delta^{(i)}(\mathbi{k}_1)
\delta^{(j)}(\mathbi{k}_2)
\delta^{(k)}(\mathbi{k}_3)
\delta^{(l)}(\mathbi{k}_4)
\rangle_c.
\label{eq:def_Tijkl}
\end{eqnarray}
The leading contributions of all the terms shown in Eq.~(\ref{eq:def_TR}) 
are order of $\phi^6$.

The first term, $T_R^{1111}$, is the linearly evolved  
primordial trispectrum calculated from Eq.~(\ref{eq:tphi}), and thus it
contains the terms proportional to $\fnl^2$ and $\gnl$.
The second term, $T_R^{1112}$, has a coupling between 
the primordial non-Gaussianity (linear in $\fnl$) and 
the non-linear gravitational evolution (linear in
$F_2^{(s)}$). 
These two terms are important on large scales.

The other terms, $T_R^{1113}$ and $T_R^{1122}$,
do not have contributions from $\fnl$ or $\gnl$ at the leading-order
level, but solely come from the non-linear gravitational coupling; thus, 
they may be ignored on large scales we are considering in this paper. 
\citet{sefusatti:prep} also derived and studied $T_R^{1112}$
as well as $T_R^{1113}$ and $T_R^{1122}$.

Therefore, we approximate the integration in 
the third term of Eq.~(\ref{eq:bisp_nG2}) as
\begin{eqnarray}
\nonumber
&&
\frac{1}{2\sigma_R^2}
\int\frac{d^3q}{(2\pi)^3}
\left[
T_R(\mathbi{q},\mathbi{k}_1-\mathbi{q},\mathbi{k}_2,\mathbi{k}_3)
+(2~\mathrm{cyclic})
\right]
\\
\nonumber
&\approx&
\frac{1}{2\sigma_R^2}
\biggl\{
\int\frac{d^3q}{(2\pi)^3}
\left[
T_R^{(1)}(\mathbi{q},\mathbi{k}_1-\mathbi{q},\mathbi{k}_2,\mathbi{k}_3)
+(2~\mathrm{cyclic})
\right]
\\
\label{eq:Tr_separate}
&&+
\int\frac{d^3q}{(2\pi)^3}
\left[
T_R^{(2)}(\mathbi{q},\mathbi{k}_1-\mathbi{q},\mathbi{k}_2,\mathbi{k}_3)
+(2~\mathrm{cyclic})
\right]\biggr\},
\end{eqnarray}
where ``cyclic'' denotes the cyclic combinations of $k_1$, $k_2$, and
$k_3$, and $T_R^{(1)}$ and $T_R^{(2)}$ denote
$
T_R^{(1)}(\mathbi{k}_1,\mathbi{k}_2,\mathbi{k}_3,\mathbi{k}_4)
=
T_R^{1111}(\mathbi{k}_1,\mathbi{k}_2,\mathbi{k}_3,\mathbi{k}_4)
$,
and 
$
T_R^{(2)}(\mathbi{k}_1,\mathbi{k}_2,\mathbi{k}_3,\mathbi{k}_4)
=
T_R^{1112}(\mathbi{k}_1,\mathbi{k}_2,\mathbi{k}_3,\mathbi{k}_4)
+(3~\mbox{cyclic}),
$ respectively.

The first term in Eq.~(\ref{eq:Tr_separate}) is the integration of the 
linearly evolved primordial curvature trispectrum, which contains two pieces:
one proportional to $\fnl^2$ and another to $\gnl$ (see Eq.~(\ref{eq:tphi})).
Therefore, we symbolically write the first line in Eq.~(\ref{eq:Tr_separate}) as
\begin{widetext}
\begin{eqnarray}
\frac{1}{2\sigma_R^2}\int\frac{d^3q}{(2\pi)^3}
\left[
T_R^{(1)}(\mathbi{q},\mathbi{k}_1-\mathbi{q},\mathbi{k}_2,\mathbi{k}_3)
+(2~\mathrm{cyclic})\right]
\label{eq:def_BnGs1}
=
\gnl B_{\gnl}^{nG}(k_1,k_2,k_3)
+
\fnl^2 B_{\fnl^2}^{nG}(k_1,k_2,k_3),
\end{eqnarray}
where
\begin{eqnarray}
\nonumber
B_{\fnl^2}^{nG}(k_1,k_2,k_3)
&\equiv &
\frac{1}{2\sigma_R^2}\biggl[
4\Mr(k_2)\Mr(k_3)
\int \frac{d^3q}{(2\pi)^3}
\Mr(q)
\Mr(|\mathbi{k}_1-\mathbi{q}|)
P_\phi(q)\\
\nonumber
& &\times
\left[
P_\phi(|\mathbi{k}_1-\mathbi{q}|)
P_\phi(|\mathbi{k}_2+\mathbi{q}|)
+
P_\phi(|\mathbi{k}_1-\mathbi{q}|)
P_\phi(|\mathbi{k}_3+\mathbi{q}|)
\right]+
(2~\mathrm{cyclic})
\\
\nonumber
&+&
8\Mr(k_2)\Mr(k_3)
P_\phi(k_2)
\int \frac{d^3q}{(2\pi)^3}
\Mr(q)
\Mr(|\mathbi{k}_1-\mathbi{q}|)
P_\phi(q)P_\phi(|\mathbi{k}_3+\mathbi{q}|)
+
(2~\mathrm{cyclic})
\\
\nonumber
&+&
8\Mr(k_2)\Mr(k_3)
P_\phi(k_3)
\int \frac{d^3q}{(2\pi)^3}
\Mr(q)
\Mr(|\mathbi{k}_1-\mathbi{q}|)
P_\phi(q)P_\phi(|\mathbi{k}_2+\mathbi{q}|)
+
(2~\mathrm{cyclic})
\\
\nonumber
&+&
8\Mr(k_2)\Mr(k_3)
P_\phi(k_1)
\left[
P_\phi(k_2)+P_\phi(k_3)
\right]
\int \frac{d^3q}{(2\pi)^3}
\Mr(q)
\Mr(|\mathbi{k}_1-\mathbi{q}|)
P_\phi(q)
+
(2~\mathrm{cyclic})
\\
\nonumber
&+&
4\Mr(k_2)\Mr(k_3)P_\phi(k_2)P_\phi(k_3)
\int \frac{d^3q}{(2\pi)^3}
\Mr(q)
\Mr(|\mathbi{k}_1-\mathbi{q}|)
\\
& &\times
\left[
P_\phi(|\mathbi{k}_2+\mathbi{q}|)
+
P_\phi(|\mathbi{k}_3+\mathbi{q}|)
\right]
+
(2~\mathrm{cyclic})\biggl],
\label{eq:bfnl2*}
\\
\nonumber
B_{\gnl}^{nG}(k_1,k_2,k_3)
& \equiv&
\frac{1}{2\sigma_R^2}\biggl[
6
\Mr(k_2)\Mr(k_3)\left[
P_\phi(k_2)+P_\phi(k_3)
\right]
\int \frac{d^3q}{(2\pi)^3}
\Mr(q)\Mr(|\mathbi{k}_1-\mathbi{q}|)
P_\phi(q)
P_\phi(|\mathbi{k}_1-\mathbi{q}|)+(2~\mathrm{cyclic})
 \\ 
&+&
12
\Mr(k_2)\Mr(k_3)P_\phi(k_2)P_\phi(k_3)
\int \frac{d^3q}{(2\pi)^3}
\Mr(q)\Mr(|\mathbi{k}_1-\mathbi{q}|)
P_\phi(q)
+(2~\mathrm{cyclic})\biggl].
\label{eq:bgnl}
\end{eqnarray}
\end{widetext}
We find that the first three cyclic terms in Eq.~(\ref{eq:bfnl2*}) are
parametrically small on large scales and may be ignored for 
$k\lesssim 0.1~h~\mathrm{Mpc}^{-1}$. Therefore, one may just
calculate the last two cyclic terms:
\begin{widetext}
\begin{eqnarray}
\nonumber
B_{\fnl^2}^{nG}(k_1,k_2,k_3)
&\approx &
\frac{1}{2\sigma_R^2}\biggl[
8\Mr(k_2)\Mr(k_3)
P_\phi(k_1)
\left[
P_\phi(k_2)+P_\phi(k_3)
\right]
\int \frac{d^3q}{(2\pi)^3}
\Mr(q)
\Mr(|\mathbi{k}_1-\mathbi{q}|)
P_\phi(q)
+
(2~\mathrm{cyclic})
\\
\nonumber
&+&
4\Mr(k_2)\Mr(k_3)P_\phi(k_2)P_\phi(k_3)
\int \frac{d^3q}{(2\pi)^3}
\Mr(q)
\Mr(|\mathbi{k}_1-\mathbi{q}|)
\\
& &\times
\left[
P_\phi(|\mathbi{k}_2+\mathbi{q}|)
+
P_\phi(|\mathbi{k}_3+\mathbi{q}|)
\right]
+
(2~\mathrm{cyclic})\biggr].
\label{eq:bfnl2}
\end{eqnarray}
\end{widetext}

Next, the second term of Eq.~(\ref{eq:Tr_separate}) contains a
cross-correlation between the  
non-linearly evolved density field ($\delta^{(2)}\sim F_2^{(s)}[\delta^{(1)}]^2$) and the primordial bispectrum, and
thus it is linearly proportional to $\fnl$ and $F_2^{(s)}$.
We present the explicit functional form of
$T_R^{1112}$ as well as the full expression of the second term of
Eq.~(\ref{eq:Tr_separate}) in Appendix \ref{sec:appA}. 
Here, we only show the final result.
We write it as
\begin{widetext}
\begin{eqnarray}
\nonumber
& &\frac{1}{2\sigma_R^2}
\int\frac{d^3q}{(2\pi)^3}
T_R^{(2)}(\mathbi{q},\mathbi{k}_1-\mathbi{q},\mathbi{k}_2,\mathbi{k}_3)
+(2~\mathrm{cyclic})\\
& &
=
\fnl
\biggl[
B_m^{nG}(k_1,k_2,k_3)
+
B_{\fnl}^{nG1}(k_1,k_2,k_3)
+
4
B_{\fnl}^{nG0}(k_1,k_2,k_3)
\left\{
\Gr(k_1)+
\Gr(k_2)+
\Gr(k_3)
\right\}
\biggl]
\label{eq:def_BnGs2},
\end{eqnarray}
where 
\begin{eqnarray}
B_m^{nG}(k_1,k_2,k_3)
&\equiv&
4
\Wr(k_1)\Wr(k_2)\Wr(k_3)
\biggl\{
\left[
\frac{\Fr(k_1)}{\Mr(k_1)}
+
\frac{\Fr(k_2)}{\Mr(k_2)}
\right]
P_m(k_1)P_m(k_2)F_2^{(s)}(\mathbi{k}_1,\mathbi{k}_2)
+(2~\mathrm{cyclic})\biggr\},
\label{eq:BmnG}
\\
\nonumber
B_{\fnl}^{nG1}(k_1,k_2,k_3)
&\equiv&
\frac{1}{2\sigma_R^2}
\biggl[
8\Wr(k_2)\Wr(k_3)
\Mm(k_3)P_m(k_2)
\\
\nonumber
&&\times
\int\frac{d^3q}{(2\pi)^3}
\Wr(|\mathbi{k}_1-\mathbi{q}|)
\Wr(q)
\Mm(|\mathbi{k}_1-\mathbi{q}|)\Mm(|\mathbi{k}_2+\mathbi{q}|)
F_2^{(s)}(-\mathbi{k}_2,\mathbi{k}_{2}+\mathbi{q})
\\
\nonumber
&&\times
\left\{
P_\phi(k_3)P_\phi(|\mathbi{k}_1-\mathbi{q}|)
+
P_\phi(k_3)P_\phi(|\mathbi{k}_{2}+\mathbi{q}|)
+
P_\phi(|\mathbi{k}_1-\mathbi{q}|)P_\phi(|\mathbi{k}_{2}+\mathbi{q}|)
\right\}
+(5~\mbox{permutation})
\\
\nonumber
&&+
8\Wr(k_2)\Wr(k_3)
\Mm(k_3)
\int\frac{d^3q}{(2\pi)^3}
\Wr(|\mathbi{k}_1-\mathbi{q}|)
\Wr(q)
\Mm(|\mathbi{k}_1-\mathbi{q}|)
\Mm(|\mathbi{k}_{2}+\mathbi{q}|)P_m(q)
F_2^{(s)}(-\mathbi{q},\mathbi{k}_{2}+\mathbi{q})
\\
\nonumber
&&\times
\left\{
P_\phi(|\mathbi{k}_1-\mathbi{q}|)P_\phi(k_3)
+
P_\phi(|\mathbi{k}_1-\mathbi{q}|)P_\phi(|\mathbi{k}_{2}+\mathbi{q}|)
+
P_\phi(k_3)P_\phi(|\mathbi{k}_{2}+\mathbi{q}|)
\right\}
+(5~\mbox{permutation})
\\
\nonumber
&&+
8\left(\Wr(k_2)\Wr(k_3)\right)^2
P_m(k_3) \Mm(k_2)
\int \frac{d^3p}{(2\pi)^3}
\Mm(p)\Mm(|\mathbi{k}_2-\mathbi{p}|)
P_\phi(p)
\\
&&\times
\left\{
P_\phi(|\mathbi{k}_2-\mathbi{p}|)
+2P_\phi(k_2)
\right\}
F_2^{(s)}(\mathbi{p},\mathbi{k}_2-\mathbi{p})
+(5~\mbox{permutation})
\biggl].
\label{eq:BfnlnG1*}
\end{eqnarray}
Here, $\Mm(k)\equiv\Mr(k)/\Wr(k)$. 
\end{widetext}

\begin{figure}
\begin{center}
\includegraphics[width=8.5cm]{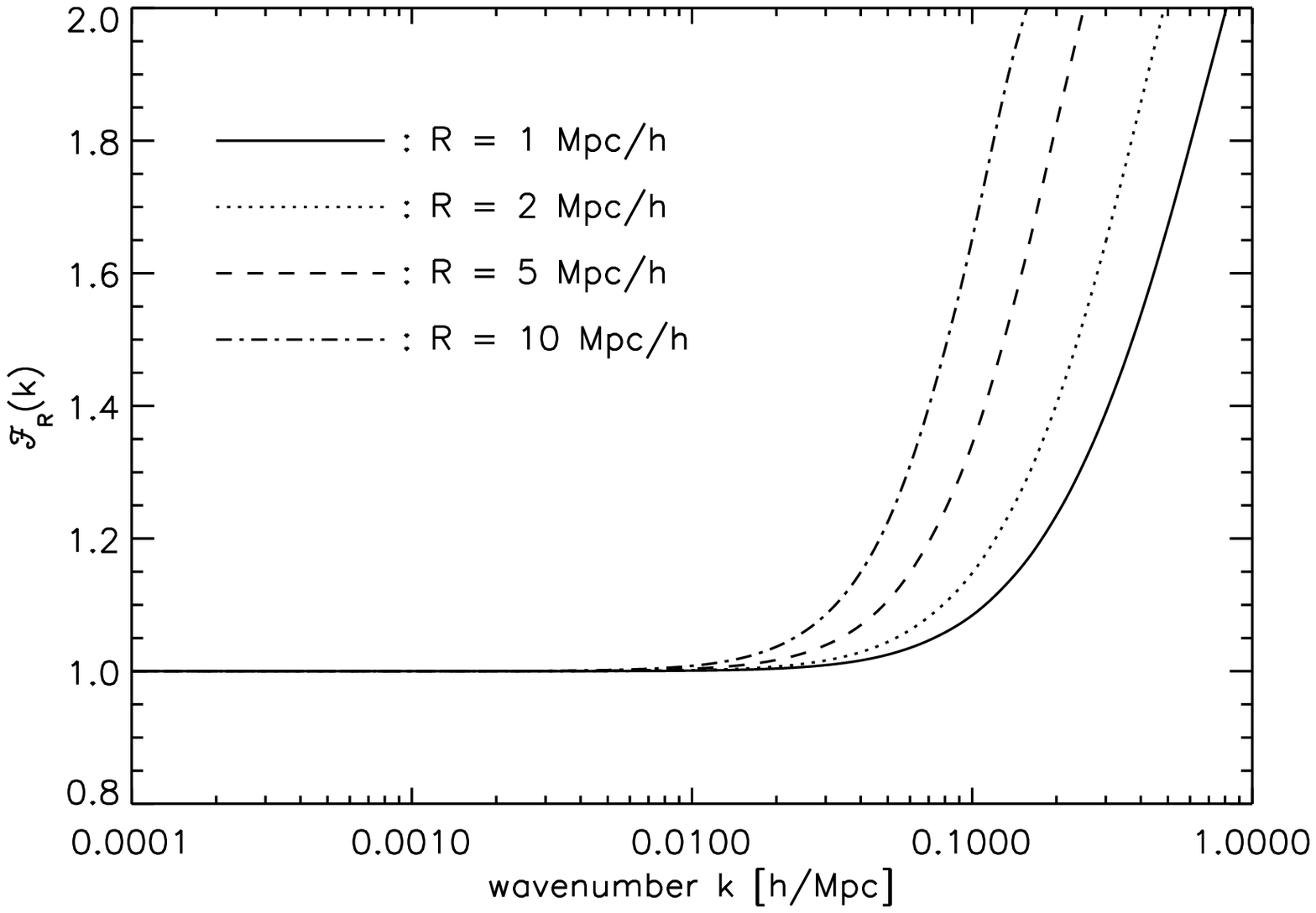}
\caption{
Shape of the function, $\Fr(k)$, defined in Eq.~(\ref{eq:def_Fr}).
We show $\Fr(k)$ for four different smoothing
lengths: $R=1$, $2$, $5$, $10~\mathrm{Mpc}/h$.
         }%
\label{fig:Fr}
\end{center}
\end{figure}

\begin{figure}
\begin{center}
\includegraphics[width=8.5cm]{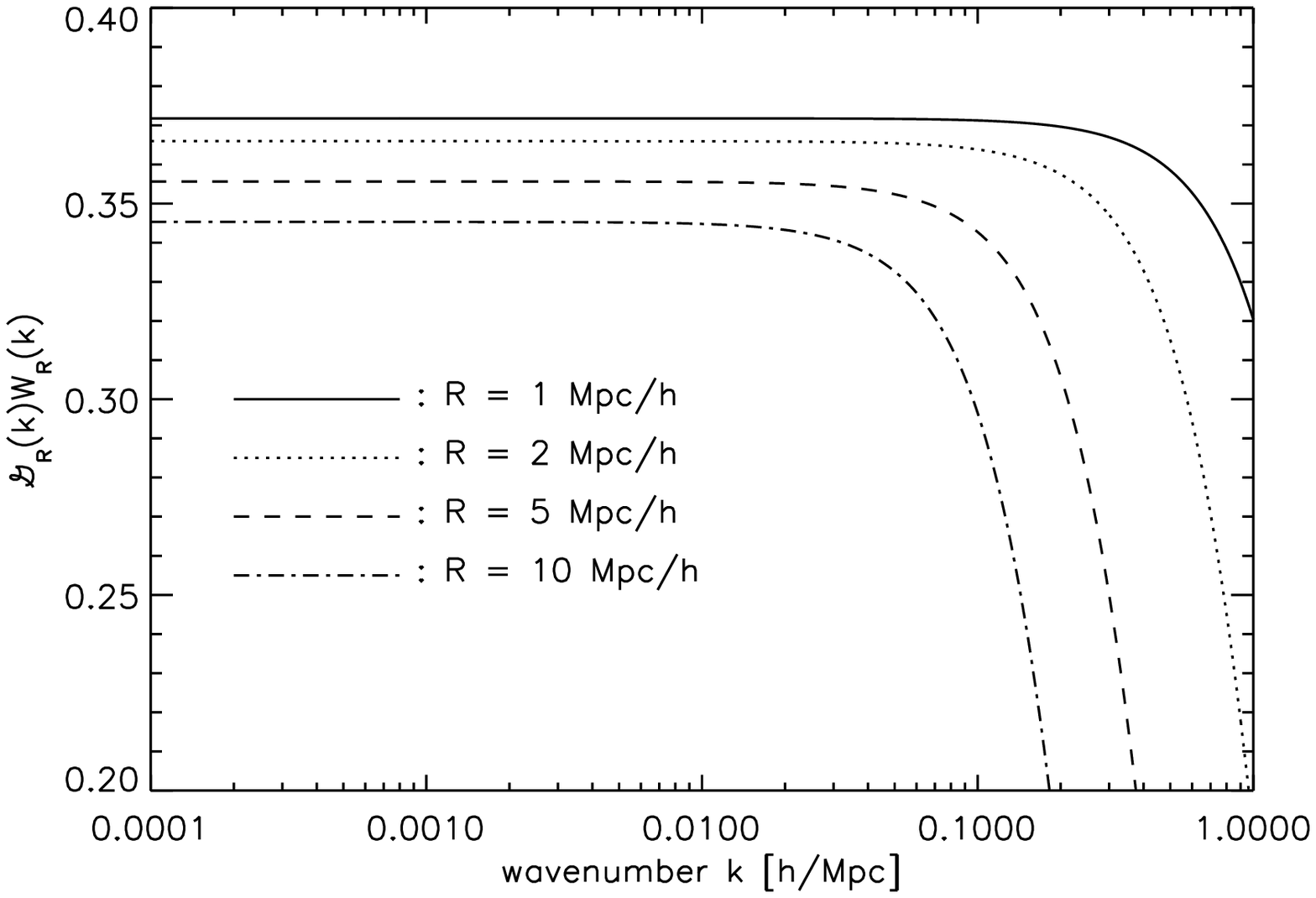}
\caption{
Shape of the function, $\Gr(k)$, defined in Eq.~(\ref{eq:def_Fr}).
We show $\Gr(k)$ for four different smoothing
lengths: $R=1$, $2$, $5$, $10~\mathrm{Mpc}/h$.
         }%
\label{fig:Gr}
\end{center}
\end{figure}

\begin{figure}
\begin{center}
\includegraphics[width=8.5cm]{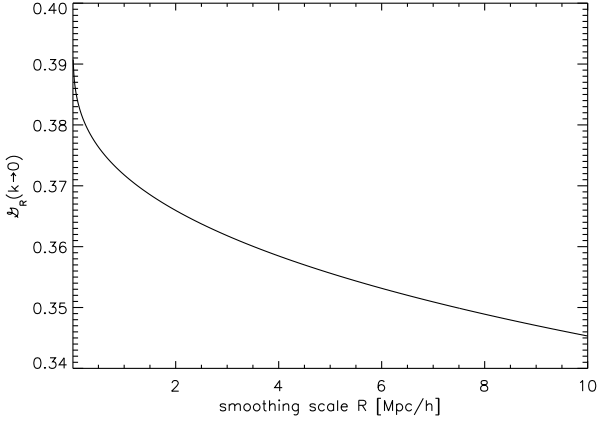}
\caption{
Large-scale asymptotic value of $\Gr(k)$ as a function of the smoothing
scale $R$. The value for $R=1~[\mathrm{Mpc}/h]$, which is used for
generating Figs.~\ref{fig:fnl_tot} to \ref{fig:z3f4}, is $0.3718$. 
         }%
\label{fig:Gr_k=0}
\end{center}
\end{figure}

In the above equations we have defined two functions, $\Fr(k)$ and
$\Gr(k)$, which are given by
\begin{eqnarray}
\nonumber
\Fr(k)
&\equiv&
\frac{1}{2\sigma_R^2}
\int\frac{d^3q}{(2\pi)^3}
P_\phi(q)
\Mr(q)
\Mr(|\mathbi{k}-\mathbi{q}|)
\\
&&\times
\left[
\frac{P_\phi(|\mathbi{k}-\mathbi{q}|)}{P_\phi(k)} +2
\right],
\label{eq:def_Fr}
\\
\nonumber
\Gr(k)
&\equiv&
\frac{1}{2\sigma_R^2}
\int \frac{d^3q}{(2\pi)^3} 
\frac
{\Wr(q)\Wr(|\mathbi{k}-\mathbi{q}|)}{\Wr(k)}
\\
&&\times
P_m(q)F_2^{(s)}(\mathbi{k},-\mathbi{q}).
\label{eq:def_Gr}
\end{eqnarray}
As shown in Fig.~\ref{fig:Fr} and Fig.~\ref{fig:Gr}, both $\Fr(k)$ and 
$\Gr(k)$ are almost constant on large scales.
If we do not have a smoothing, i.e., $R\to 0$, 
the large scale asymptotic value of $\Gr(k)$ 
is $17/42$. However, the presence of filter changes this asymptotic
value. As $k\to0$, 
\begin{equation}
\Gr(k)\to \frac{13}{84} + \frac{1}{4\sigma_R^2}
\int\frac{d^3q}{(2\pi)^3} \Wr(q)P_m(q)\frac{\sin(qR)}{qR},
\end{equation}
whose value depends on the smoothing scale, $R$, as shown in 
Fig.~\ref{fig:Gr_k=0}.

Let us study the structure of each term in Eq.~(\ref{eq:def_BnGs2}).
The first piece is $B_{m}^{nG}$.
On very large scales, where $\Wr(k)\to1$ and $\Fr(k)\to1$, 
$B_{m}^{nG}$ becomes a  product of 
the usual matter bispectrum for Gaussian initial conditions, $B_m^G$,
and the scale dependent bias shown in Eq.~(\ref{eq:bk}), as
\begin{equation}
\frac{2\fnl\Fr(k)}{\Mr(k)}
=
\frac{3\fnl H_0^2\Omega_m}{k^2T(k)}\frac{\Fr(k)}{\Wr(k)}
\to
\frac{3\fnl H_0^2\Omega_m}{k^2T(k)},
\end{equation}
as $k\to 0$.
Therefore, we can interpret this term as a scale dependent bias
multiplying the usual matter bispectrum for Gaussian initial conditions;
however, this behavior is not generic -- in fact, the other terms cannot
be expressed in terms of products of the scale-dependent bias and the
results in the continuous limit, Eq.~(\ref{eq:SK07}).
 
The next piece is $B_{\fnl}^{nG1}(k_1,k_2,k_3)$.
By numerically calculating Eq.~(\ref{eq:BfnlnG1*}), 
we find that the terms that contain
$F_2^{(s)}(\mathbi{q},\mathbi{k}-\mathbi{q})$ 
are parametrically small on large scales, and that 
the dominant contributions come from the first permutation terms.
Therefore, we approximate Eq.~(\ref{eq:BfnlnG1*})
on large scale ($k\lesssim 0.1 h \mathrm{Mpc}^{-1}$) as
\begin{widetext}
\begin{eqnarray}
\nonumber
B_{\fnl}^{nG1}(k_1,k_2,k_3)
&\approx&
\frac{1}{2\sigma_R^2}
\biggl[
8\Wr(k_2)\Wr(k_3)
P_m(k_2)\Mm(k_3)P_\phi(k_3)
\int \frac{d^3q}{(2\pi)^3} 
\Wr(|\mathbi{k}_1-\mathbi{q}|)
\Wr(q)
\Mm(|\mathbi{k}_1-\mathbi{q}|)
\Mm(|\mathbi{k}_2+\mathbi{q}|)
\\
&&\times
\left[
P_\phi(|\mathbi{k}_2+\mathbi{q}|) + 
P_\phi(|\mathbi{k}_1-\mathbi{q}|)
\right]
F_2^{(s)}(-\mathbi{k}_2,\mathbi{k}_2+\mathbi{q})
+(5~\mathrm{permutation})\bigg].
\label{eq:BfnlnG1}
\end{eqnarray}
\end{widetext}

How about the last term of Eq.~(\ref{eq:def_BnGs2}),
$4B_{\fnl}^{nG0}(k_1,k_2,k_3)
\left\{
\Gr(k_1)+
\Gr(k_2)+
\Gr(k_3)\right\}$? 
As $\Gr(k)\to {\rm constant}$ on large scales (Fig.~\ref{fig:Gr}), 
this piece becomes $B_{\fnl}^{nG0}$ multiplied by a constant factor
whose exact value depends on 
the smoothing scale, $R$ (Fig.~\ref{fig:Gr_k=0}).

In summary, we have derived the new terms in the galaxy bispectrum, 
which arise from the integration of the matter trispectrum.
While we find one term, $B_m^{nG}$, includes the scale-dependent bias
 which appears on the galaxy power spectrum, we also find that 
there are more terms contributing to the galaxy bispectrum.


Eq.~(\ref{eq:def_BnGs1}) along with 
Eqs.~(\ref{eq:bfnl2*})--(\ref{eq:bfnl2}), and
Eq.~(\ref{eq:def_BnGs2}) along with 
Eqs.~(\ref{eq:BmnG}), (\ref{eq:BfnlnG1*}), (\ref{eq:BfnlnG1})
are the second main results of this paper.
In the next sections, we shall present the detailed assessment of 
the new terms we have derived in this section.

\subsubsection{Shape Dependence}
Let us consider the shape dependence.
First of all, the last term of Eq.~(\ref{eq:def_BnGs2}) has the same shape 
dependence as $B_{\fnl}^{nG0}$, as $\Gr(k)$ is almost constant on large scale.
Thus, it peaks at the squeezed configurations as $B_{\fnl}^{nG0}$ does.
How about the shape dependence of the other terms?

All terms in Eqs.~(\ref{eq:bgnl}), (\ref{eq:bfnl2}), (\ref{eq:BfnlnG1}) 
have $P_\phi(k_i)$ outside of the integral, 
and Eq.~(\ref{eq:BmnG}) contains  $1/\Mr(k)\propto k^{-2}$,
which suggests that 
all of $B_{\gnl}^{nG}$, $B_{\fnl^2}^{nG}$, $B_{\fnl}^{nG1}$, and
$B_m^{nG}$ peak at the squeezed configurations. 
For sufficiently large scales in which $T(k)\approx 1$, and 
for a scale-invariant spectrum ($P_\phi(k)\propto 1/k^3$), we may write
down Eqs.~(\ref{eq:BmnG}), (\ref{eq:BfnlnG1}), (\ref{eq:bfnl2}), and
(\ref{eq:bgnl}),  as
\begin{widetext}
\begin{eqnarray}
 B_m^{nG}(k_1,k_2,k_3)
&\propto&
\left(\frac{k_2}{k_1}+\frac{k_1}{k_2}\right)
\left[
\frac{5}{7} + 
\frac{\mathbi{k}_1\cdot\mathbi{k}_2}{2k_1k_2}\left(\frac{k_1}{k_2}+\frac{k_2}{k_1}\right)
+\frac{2}{7}
\left(\frac{\mathbi{k}_1\cdot\mathbi{k}_2}{k_1k_2}\right)^2
\right]
+(2~\mbox{cyclic}),
\label{eq:bmngapp}
\\
\nonumber
 B_{\fnl}^{nG1}(k_1,k_2,k_3)
&\propto&
\frac{k_2}{k_3}
\int \frac{d^3q}{q}
q^3 
\left(
\frac{|\mathbi{k}_1-\mathbi{q}|^2}{|\mathbi{k}_2+\mathbi{q}|}
+
\frac{|\mathbi{k}_2+\mathbi{q}|^2}{|\mathbi{k}_1-\mathbi{q}|}
\right)
\Wr(|\mathbi{k}_1-\mathbi{q}|)\Wr(q)
T(|\mathbi{k}_1-\mathbi{q}|)T(|\mathbi{k}_2+\mathbi{q}|)
\\
&&\times
F_2^{(s)}(-\mathbi{k}_2,\mathbi{k}_2+\mathbi{q})
+(5~\mbox{permutation}),
\label{eq:bfnl1app}
\\
\nonumber
 B_{\fnl^2}^{nG}(k_1,k_2,k_3)
&\propto&
\frac{4}{k_2k_3}\left[
\frac{2(k_2^3+k_3^3)}{k_1^3}
\int \frac{d^3q}{q} |{\mathbi k}_1-\mathbi{q}|^2
T(q)\tilde{W}_R(q)
T(|{\mathbi k}_1-\mathbi{q}|)
\tilde{W}_R(|{\mathbi k}_1-\mathbi{q}|)
\right.\\
& &
\left.
+
\int d^3q~q^2|{\mathbi k}_1-\mathbi{q}|^2\left(\frac1{|{\mathbi k}_2+\mathbi{q}|^3}+\frac1{|{\mathbi k}_3+\mathbi{q}|^3}\right)
T(q)\tilde{W}_R(q)
T(|{\mathbi k}_1-\mathbi{q}|)
\tilde{W}_R(|{\mathbi k}_1-\mathbi{q}|)
\right]+(2~\mbox{cyclic}),
\label{eq:bfnl2app}
\\
\nonumber
 B_{\gnl}^{nG}(k_1,k_2,k_3)
&\propto&
\frac{6}{k_2k_3}\left[
(k_2^3+k_3^3)
\int \frac{d^3q}{q} \frac{1}{|{\mathbi k}_1-\mathbi{q}|}
T(q)\tilde{W}_R(q)
T(|{\mathbi k}_1-\mathbi{q}|)
\tilde{W}_R(|{\mathbi k}_1-\mathbi{q}|)
\right.\\
& &
\left.
+
2\int \frac{d^3q}{q}|{\mathbi k}_1-\mathbi{q}|^2
T(q)\tilde{W}_R(q)
T(|{\mathbi k}_1-\mathbi{q}|)
\tilde{W}_R(|{\mathbi k}_1-\mathbi{q}|)
\right]+(2~\mbox{cyclic}),
\label{eq:bgnlapp}
\end{eqnarray}
respectively.
\end{widetext}
For a given $k_1$, all of these terms have the largest signals when 
$k_3$ is small, i.e., the squeezed configurations.
Note that we do not use the exact scale-invariant spectrum for the
numerical calculation, but use the WMAP 5-year best-fitting value
reported in Table 1 (``WMAP+BAO+SN'') of
\citet{komatsu/etal:2009}. 

The top-left and bottom-left panels of 
Fig.~\ref{fig:bkNG_fnl} show 
$B_{m}^{nG}$ and $B_{\fnl}^{nG1}$ as a function 
of $k_2/k_1$ and $k_3/k_1$, respectively,
 for $k_1=0.01~h~{\rm Mpc}^{-1}$.
The top-right and bottom-right panels of 
Fig.~\ref{fig:bkNG_fnl} show the same quantities 
for $k_1=0.05~h~{\rm Mpc}^{-1}$.
We also show $B_{\gnl}^{nG}$ and $B_{\fnl^2}^{nG}$ in the 
top-left and bottom-left panels of Fig.~\ref{fig:bkNG} 
for $k_1=0.01~h~{\rm Mpc}^{-1}$, and top-right and bottom-right for
for $k_1=0.05~h~{\rm Mpc}^{-1}$.
In all cases we find that $B_m^{nG}$, $B_{\fnl}^{nG1}$, 
$B_{\gnl}^{nG}$ and $B_{\fnl^2}^{nG}$
peak at the squeezed configurations, as expected from the above
argument. 

We find that the shape dependence of $B_{\fnl}^{nG0}$ and that of
$B_{\fnl}^{nG1}$, $B_{\gnl}^{nG}$ are quite similar, whereas 
that of $B_{m}^{nG}$ is higher toward the elongated triangles, and
that of $B_{\fnl^2}^{nG}$ is more sharply peaked at the 
squeezed configuration. 

We can understand this behavior analytically as follows. In order to
simplify the analysis, we consider a scale-invariant curvature power spectrum, 
$P_\phi = P_{\phi 0}/k^3$,
on large scales where Eqs.~(\ref{eq:bfnl1app}), (\ref{eq:bmngapp}), 
(\ref{eq:bfnl2app}), and (\ref{eq:bgnlapp}) are valid.
On such a large scale, $\Mr(k)$ can be approximated as
$\Mr(k) \simeq 2k^2/(3H_0^2\Omega_m) \equiv \M0 k^2$,
where 
$M_0\simeq2.16\times10^7\left(0.277/\Omega_m\right) [\mathrm{Mpc}/h]^2$ 
is a constant.
We focus on the squeezed triangle, $k_1=k_2=\alpha k_3\equiv k$
($\alpha\gg1$), where the signals of all the primordial non-Gaussianity
terms are maximized. 
The triangles in this configuration lie on the upper side of 
the triangular region of $(k_3/k_1,k_2/k_1)$ plane in 
Fig.~\ref{fig:bkNG_fnl} and Fig.~\ref{fig:bkNG}, and
the triangle approaches the exact squeezed limit as $\alpha\to\infty$.
With this parametrization, we compare the dominant contributions 
of each of these primordial non-Gaussianity terms. 

First, we shall analyze the terms proportional to $\fnl$: 
$B_{\fnl}^{nG0}$, $B_m^{nG}$, and $B_{\fnl}^{nG1}$.
The largest contribution to $B_{\fnl}^{nG0}$ in the squeezed
configurations occurs when $k_3$ is in the denominator:
\begin{eqnarray}
\nonumber
B_{\fnl}^{nG0} 
&=& 
2\M0^3P^2_{\phi 0}
\left(\frac{k_1^2}{k_2k_3}+\frac{k_2^2}{k_3k_1}+
\frac{k_3^2}{k_1k_2}
\right)
\\
\nonumber
&\simeq&
2\M0^3P^2_{\phi 0}
\left(
\frac{k_1^2}{k_2k_3}+\frac{k_2^2}{k_3k_1}
\right)
\\
\label{eq:bfnl_squeeze}
&=&
4\alpha\M0^3
P^2_{\phi 0}.
\end{eqnarray}

To compute $B_m^{nG}$, which contains $F_2^{(s)}$, we note that, 
in the squeezed limit, the angular cosines between two wave vectors are
$\mathbi{k}_1\cdot\mathbi{k}_2/(k_1k_2) = -1+1/\alpha^2 \simeq -1$
and 
$\mathbi{k}_2\cdot\mathbi{k}_3/(k_2k_3) = 
 \mathbi{k}_1\cdot\mathbi{k}_3/(k_1k_3) = -1/(2\alpha)$. We thus find
\begin{eqnarray}
\nonumber
B_m^{nG}
&=&
8\M0^3P^2_{\phi 0}
\left(\alpha+\frac{1}{\alpha}\right)
\left[
\frac{5}{7} -\frac{1}{4\alpha}\left(\alpha+\frac{1}{\alpha}\right)
+\frac{1}{14\alpha^2}
\right]
\\
&\simeq&
\frac{26}{7}\alpha
\M0^3P^2_{\phi 0}.
\label{eq:bmnG_squeeze}
\end{eqnarray}

The detailed analysis for $B_{\fnl}^{nG1}$ is more complicated, as 
Eq.~(\ref{eq:BfnlnG1}) involves a non-trivial integration. 
We simplify the situation by only analyzing the dominant term, which can
be written as
\begin{eqnarray}
\nonumber
B_{\fnl}^{nG1}
&\approx&
8\M0^3P^2_{\phi 0}
\left[
\frac{k_2}{k_3}\mathcal{H}(\mathbi{k}_1,\mathbi{k}_2)+
\frac{k_1}{k_3}\mathcal{H}(\mathbi{k}_2,\mathbi{k}_1)
\right]
\\
&=&
8\alpha \M0^3P^2_{\phi 0}
\left[
\mathcal{H}(\mathbi{k}_1,\mathbi{k}_2)+
\mathcal{H}(\mathbi{k}_2,\mathbi{k}_1)
\right],
\label{eq:bfnl1_squeeze}
\end{eqnarray}
where $\mathcal{H}(\mathbi{k}_1,\mathbi{k}_2)$ is the integration that appears
in Eq.~(\ref{eq:BfnlnG1}) including $1/(2\sigma_R^2)$ pre-factor. 
Note that this integration depends only on the magnitudes of two vectors
and  the angle between them; thus, 
for the squeezed configuration we are interested in here, 
$\mathcal{H}(\mathbi{k}_2,\mathbi{k}_1)$ depends only weakly 
on $\alpha$ -- they depend on $\alpha$ only through the inner product of 
$\mathbi{k}_1\cdot\mathbi{k}_2=k^2(-1+1/\alpha^2)$.

\begin{figure}
\begin{center}
\includegraphics[width=8.5cm]{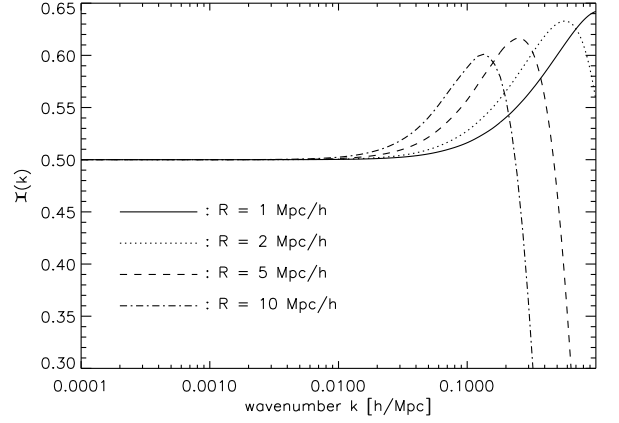}
\caption{
Shape of the integration that appears in the dominant term of 
$B_{\fnl^2}^{nG}$ and $B_{\gnl}^{nG}$, Eq.~(\ref{eq:int}).
We use four different smoothing scales: 
$R=1~$, $2$, $5$, and $10~\mathrm{Mpc}/h$. 
         }%
\label{fig:int}
\end{center}
\end{figure}

Second, we analyze $B_{\gnl}^{nG}$. 
We find that the first cyclic terms in Eq.~(\ref{eq:bgnl}) are
small in the squeezed limit, and 
the dominant contribution to $B_{\gnl}^{nG}$ is given by
\begin{eqnarray}
\nonumber
B_{\gnl}^{nG}
&=&
12
\M0^2P^2_{\phi 0}
\left[\frac{\mathcal{I}(k_1)}{k_2k_3}+\frac{\mathcal{I}(k_2)}{k_3k_1}+\frac{\mathcal{I}(k_3)}{k_1k_2}\right]
\\
&\simeq&
12
\M0^2P^2_{\phi 0}
\left[\frac{\mathcal{I}(k_1)}{k_2k_3}+\frac{\mathcal{I}(k_2)}{k_3k_1}\right],
\end{eqnarray}
where we have defined 
\begin{equation}
 \mathcal{I}(k)
\equiv
\frac{1}{2\sigma_R^2}
\int\frac{d^3q}{(2\pi)^3}
\Mr(q)\Mr(|\mathbi{k}-\mathbi{q}|)P_{\phi}(q).
\label{eq:int}
\end{equation}
We find that $\mathcal{I}(k)\simeq 0.5$ and is almost independent of $k$
on large scales (e.g., $k\lesssim 0.03~h~{\rm Mpc}^{-1}$ for 
$R=1.0~{\rm Mpc}/h$; see Fig.~\ref{fig:int}).
Therefore, by writing $\mathcal{I}(k)=\mathcal{I}_0$, we obtain
\begin{eqnarray}
B_{\gnl}^{nG}
\label{eq:bgnl_squeeze}
&\simeq& 
24\alpha
\M0^2P^2_{\phi 0} 
\frac{\mathcal{I}_0}{k^2}.
\end{eqnarray}

These results show that all the terms we have analyzed analytically so far, 
$B_{\fnl}^{nG0}$, $B_{m}^{nG}$, $B_{\fnl}^{nG1}$, and $B_{\gnl}^{nG}$,
have the same shape (i.e., $\alpha$) dependence in the squeezed configurations: 
they both increase linearly as $\alpha$ increases. This explains the
shape dependence computed from the full numerical calculations presented in
Fig.~\ref{fig:bkNG_fnl}
and the top panels of Fig.~\ref{fig:bkNG}.

\begin{figure*}[ht]
\begin{center}
\rotatebox{90}{%
\includegraphics[width=12.5cm]{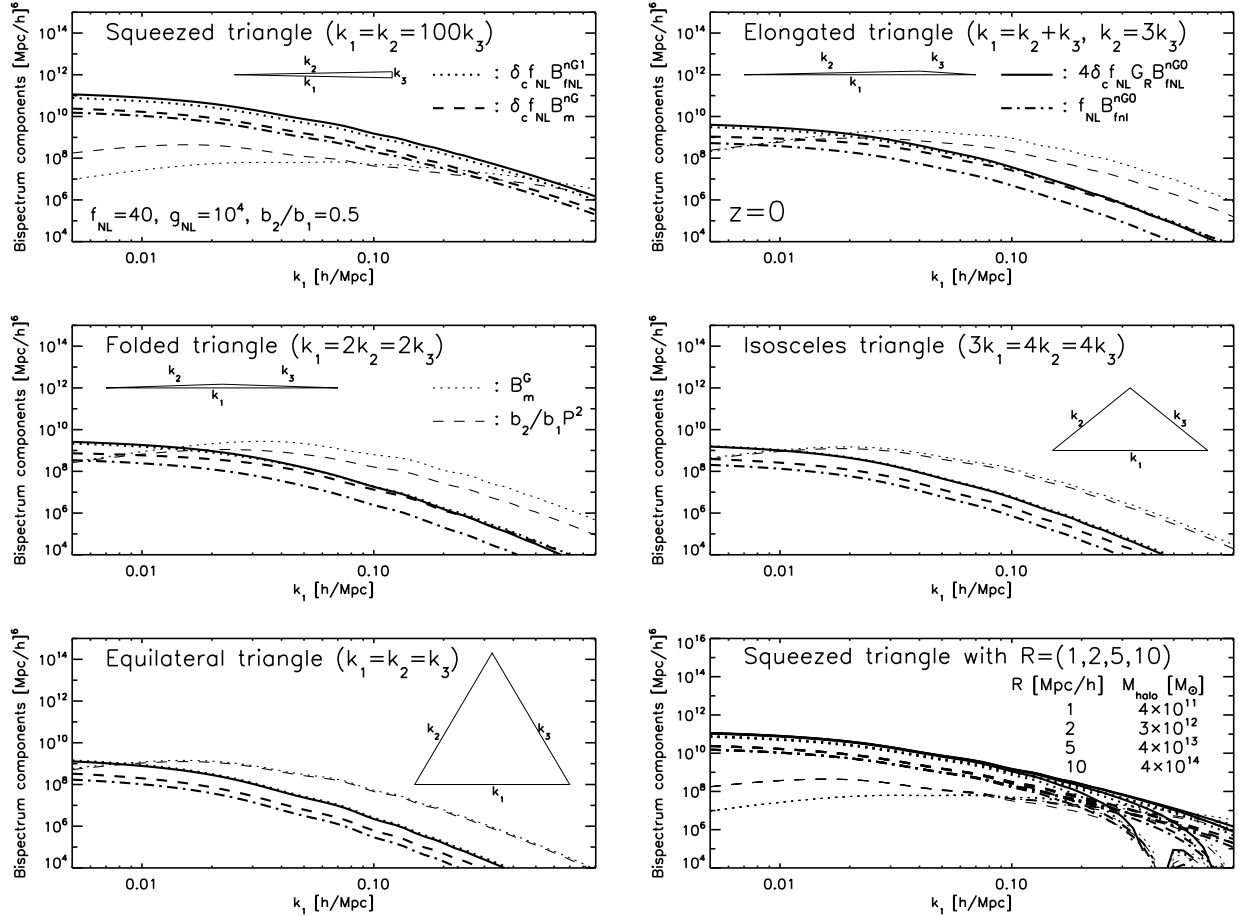}
}%
\caption{
 Scale and shape dependence of the galaxy bispectrum terms that are
 linearly proportional to $\fnl$, as a function of $k_1$.
 Except for the bottom-right panel, we use $R=1~\mathrm{Mpc}/h$.
(Top Left) The squeezed triangles with $k_1=k_2=100k_3$, (Top Right) the
 elongated triangles with
 $k_1=k_2+k_3$ and $k_2=3k_3$, (Middle Left) the folded triangles with
 $k_1=2k_2=2k_3$, (Middle Right) the isosceles triangles with
 $3k_1=4k_2=4k_3$, and (Bottom Left) the equilateral triangles with 
 $k_1=k_2=k_3$. The thick dot-dashed, dashed, solid, and dotted lines
 show the contributions from the primordial non-Gaussianity: 
 the $B_{\fnl}^{nG0}$ (Eq.~(\ref{eq:bfnl})), $\tilde{b}_2/b_1B_{m}^{nG}$ 
 (Eq.~(\ref{eq:BmnG})),
 $4(\tilde{b}_2/b_1)\left[\Gr(k_1)+\Gr(k_2)+\Gr(k_3)\right]B_{\fnl}^{nG0}$ 
 ($\Gr(k)$ defined in Eq.~(\ref{eq:def_Gr})), and $\tilde{b}_2/b_1B_{\fnl}^{nG1}$
 (Eq.~(\ref{eq:BfnlnG1})) terms, respectively.
 The thin dotted and dashed lines show the non-linear effects: 
 $B_m^G$ (Eq.~(\ref{eq:bmg})) and the
 non-linear bias (the second term in Eq.~(\ref{eq:bisp_nG})), respectively.
 We use the standard value of $\tilde{b}_2/b_1\equiv \delta_c\simeq 1.686$ from spherical collapse model.
 (Bottom Right) Dependence of the squeezed bispectrum on
 the smoothing scale, $R$, showing that the dependence is negligible for
 $k_1\ll 1/R$. 
         }%
\label{fig:fnl_tot}
\end{center}
\end{figure*}
\begin{figure*}[ht]
\begin{center}
\rotatebox{90}{%
\includegraphics[width=12.5cm]{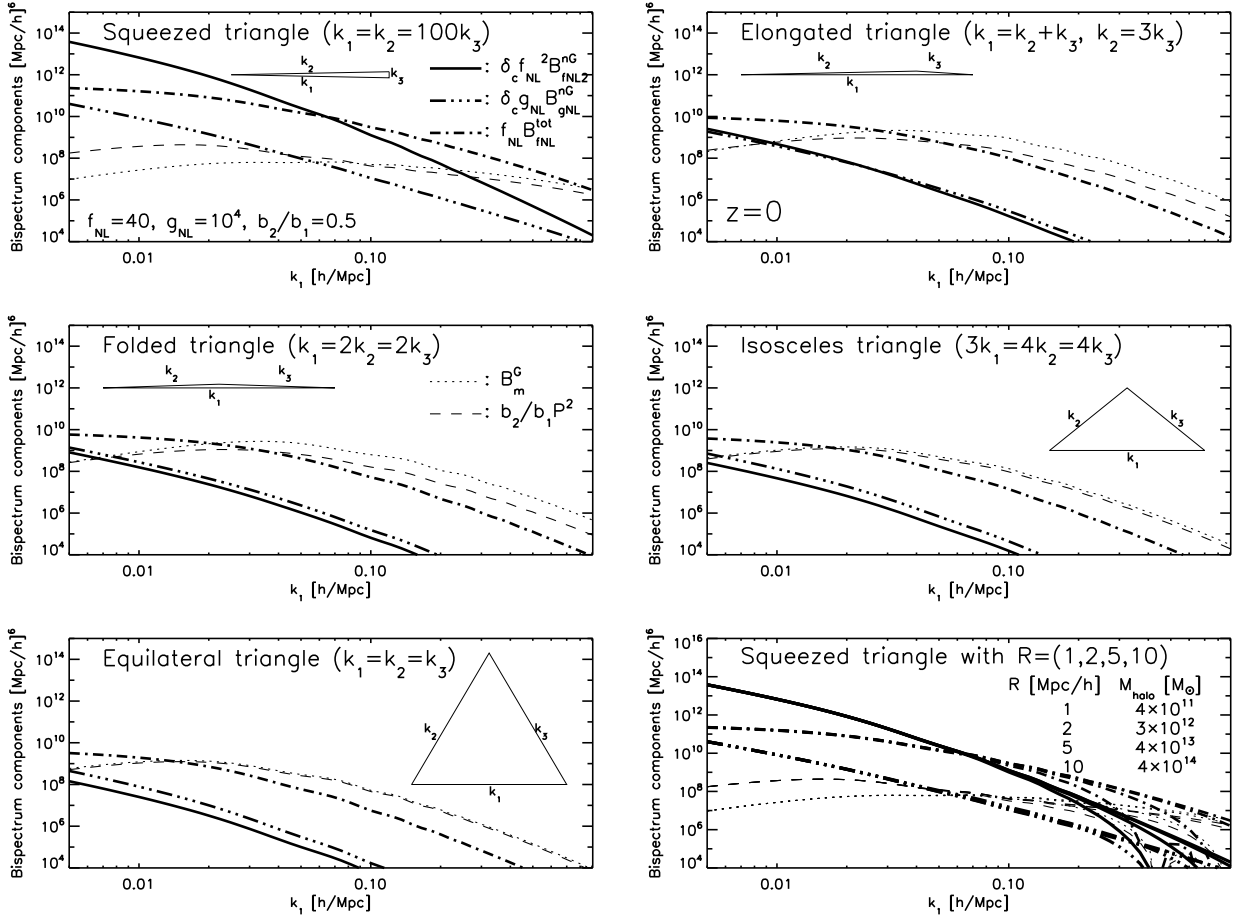}
}%
\caption{
 Scale and shape dependence of various bispectrum terms,
 $B(k_1,k_2,k_3)$, as a function of $k_1$.
 For the figure except for the bottom right, we use $R=1~\mathrm{Mpc}/h$.
(Top Left) The squeezed triangles with $k_1=k_2=100k_3$, (Top Right) the
 elongated triangles with
 $k_1=k_2+k_3$ and $k_2=3k_3$, (Middle Left) the folded triangles with
 $k_1=2k_2=2k_3$, (Middle Right) the isosceles triangles with
 $3k_1=4k_2=4k_3$, and (Bottom Left) the equilateral triangles with 
 $k_1=k_2=k_3$. The thick dot-dashed, triple-dot-dashed, and solid lines
 show the contributions from the primordial non-Gaussianity: the 
 $\fnl B_{\fnl}^{tot}$
  (Eq.~(\ref{eq:def_Bfnl_tot})), 
 $\tilde{b}_2/b_1 \gnl B_{\gnl}^{nG}$ (Eq.~(\ref{eq:bgnl})), and 
 $\tilde{b}_2/b_1 \fnl^2 B_{\fnl^2}^{nG}$
 (Eq.~(\ref{eq:bfnl2})) terms, respectively. The thin dotted and dashed
 lines show the non-linear effects: $B_m^G$ (Eq.~(\ref{eq:bmg})) and the
 non-linear bias (the second term in Eq.~(\ref{eq:bisp_nG})), respectively.
 We use the standard value of $\tilde{b}_2/b_1\equiv \delta_c\simeq 1.686$ from spherical collapse model.
 (Bottom Right) Dependence of the squeezed bispectrum on
 the smoothing scale, $R$, showing that the dependence is negligible for
 $k_1\ll 1/R$. 
         }%
\label{fig:z0f40}
\end{center}
\end{figure*}
\begin{figure*}
\begin{center}
\rotatebox{90}{%
\includegraphics[width=12.5cm]{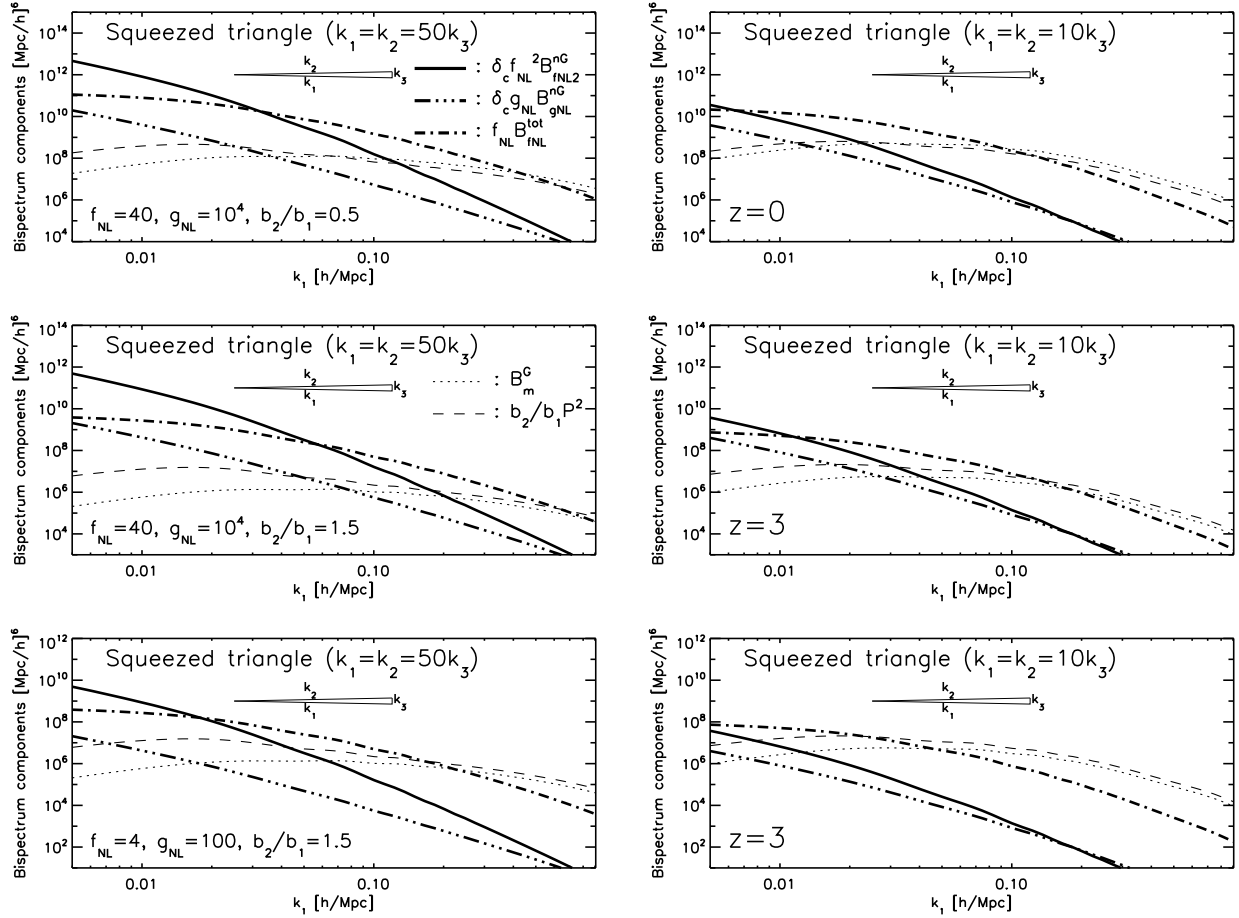}
}%
\caption{
 Same as Fig.~\ref{fig:z0f40}, but for squeezed triangles with different
 ratios: $\alpha=50$ and $\alpha=10$.
 (Top) All the parameters are the same as in Fig.~\ref{fig:z0f40}.
 (Middle) $z=3$ and $b_2/b_1=1.5$. The non-Gaussianity parameters, 
 $\fnl=40$ and $\gnl=10^4$,  are the same as in Fig.~\ref{fig:z0f40}.
 (Bottom) $z=3$ and $b_2/b_1=1.5$. The non-Gaussianity parameters, 
 $\fnl=4$ and $\gnl=100$. 
         }%
\label{fig:mild_squeeze}
\end{center}
\end{figure*}

Finally, we analyze $B_{\fnl^2}^{nG}$. 
We find that the second cyclic terms in 
Eq.~(\ref{eq:bfnl2}) are small in the squeezed configurations.
The dominant terms are:
\begin{eqnarray}
\nonumber
B_{\fnl^2}^{nG}
&=&
8\M0^2P^2_{\phi 0} \\
\nonumber
& &\times
\left[
\frac{k_2^3+k_3^3}{k_2k_3k_1^3}\mathcal{I}(k_1)+
\frac{k_3^3+k_1^3}{k_3k_1k_2^3}\mathcal{I}(k_2)+
\frac{k_1^3+k_2^3}{k_1k_2k_3^3}\mathcal{I}(k_3)\right]
\\
\nonumber
&\simeq&
8\M0^2P^2_{\phi 0} 
\frac{k_1^3+k_2^3}{k_1k_2k_3^3}\mathcal{I}(k_3)
\\
\label{eq:bfnl2_squeeze}
&\simeq&
16\alpha^3
\M0^2P^2_{\phi 0} 
\frac{\mathcal{I}_0}{k^2}.
\end{eqnarray}
Therefore, $B_{\fnl^2}^{nG}$ increases more sharply as it
approaches the squeezed limit, $B_{\fnl^2}^{nG}\propto\alpha^3$.

This sharp increase of $B_{\fnl^2}^{nG}$ relative to the other terms,
and that there are many new terms that are of the same order of magnitude as
$B_{\fnl}^{nG0}$, 
imply that the formula derived by \citet{sefusatti/komatsu:2007}
may not be valid in the squeezed configuration, where  $B_{\fnl^2}^{nG}$
may dominate over $B_{\fnl}^{nG}$. This is particularly important
because it is the squeezed configuration that gives the largest
signal from the primordial non-Gaussianity. We shall study this point in more
detail in the next section.

A careful inspection of Eq.~(\ref{eq:bfnl2app}) shows that the second term
within the square bracket diverges when $\mathbi{k}_2+\mathbi{q}=0$ or 
$\mathbi{k}_3+\mathbi{q}=0$. This is due to the fact that
$P_\phi(k)\propto 1/k^{4-n_s}$ and thus $P_\phi(k)$ diverges as $k\to 0$
for $n_s<4$. To avoid the divergence we set $P_\phi(k)=0$ at $k\le
k_{min}$, and use $k_{min}=10^{-6}~h~{\rm Mpc}^{-1}$. Fortunately the
divergence is mild and the results on the squeezed configurations, 
for which $B_{\fnl^2}^{nG}$ gives the most important contribution, 
are insensitive to $k_{min}$: changing $k_{min}=10^{-6}~h~{\rm Mpc}^{-1}$ to 
$k_{min}=10^{-9}~h~{\rm Mpc}^{-1}$ results in negligible changes in the
squeezed configurations. 

On the other hand, the folded and equilateral configurations are
more sensitive to $k_{min}$, and we find that the difference between 
$k_{min}=10^{-6}~h~{\rm Mpc}^{-1}$ and
$k_{min}=10^{-9}~h~{\rm Mpc}^{-1}$ is scale-dependent: 
at $k_1=0.01~h~{\rm Mpc}^{-1}$ the differences are negligible for
all shapes, whereas the differences reach $\sim 40$\% at $k\sim 1~h~{\rm
Mpc}^{-1}$.
(Note that the difference in the squeezed configuration reaches 1\% at
$k\sim 1~h~{\rm Mpc}^{-1}$, being totally negligible on larger scales.)
While this divergence does not have much observational consequences
(because the signals in the other configurations at $k\gtrsim 0.01~h~{\rm
Mpc}^{-1}$ would be dominated by
the other non-linear effects: $B_m^G$, non-linear bias and 
terms proportional to $\fnl$, as we show in the next section), 
there may be a better treatment of the divergence than setting 
$P_\phi(k)=0$ at $k\le k_{min}$.

\subsection{Scale Dependence}

How important are the primordial non-Gaussianity terms,
$B_{\fnl}^{nG0}$, $B_{\fnl}^{nG1}$, $B_m^{nG}$, $B_{\fnl^2}^{nG}$, 
and $B_{\gnl}^{nG}$, relative to $B_m^G$ and the non-linear bias term? 
Which one is the most dominant of the primordial terms,
terms proportional to $\fnl$, $B_{\fnl^2}^{nG}$, or $B_{\gnl}^{nG}$?
How about the scale-dependence? How about the shape dependence? 

We collect all the terms proportional to $\fnl$, and call it
$B_{\fnl}^{tot}$:
\begin{eqnarray}
\nonumber
B_{\fnl}^{tot}
&\equiv&
B_{\fnl}^{nG0} + 
\frac{\tilde{b}_2}{b_1}
\left[
B_m^{nG} 
+B_{\fnl}^{nG1}
\right.
\\
&&+ 
\left.
4\left(\Gr(k_1)+\Gr(k_2)+\Gr(k_3)\right)B_{\fnl}^{nG0}
\right].
\label{eq:def_Bfnl_tot}
\end{eqnarray}
Throughout this section, we use the standard value of 
$\tilde{b}_2/b_1=3(12\pi)^{2/3}/20\simeq 1.68$ from a spherical 
collapse model.

Figure~\ref{fig:fnl_tot} shows the scale and shape dependence of each term
in Eq.~(\ref{eq:def_Bfnl_tot}) evaluated at $z=0$. 
For all configurations shown in this figure, the primordial non-Gaussian
term calculated in 
\citet{sefusatti/komatsu:2007} is the smallest among four $\fnl$ terms, which
means that the non-Gaussian signal on large scales
is much bigger than recognized before.

For the squeezed triangle, all of the terms in
Eq.~(\ref{eq:def_Bfnl_tot}) 
depend on $k_1$ in a similar way. 
We find their ratios by comparing Eqs.~(\ref{eq:bfnl_squeeze}),
(\ref{eq:bmnG_squeeze}), and (\ref{eq:bfnl1_squeeze}):
\begin{equation}
B_{\fnl}^{nG0}:B_{m}^{nG}:B_{\fnl}^{nG1}
\simeq
1 : \frac{26}{28} : 2.96.
\end{equation}
Note that we have used the numerical value of 
$\mathcal{H}(\mathbi{k}_1,\mathbi{k}_2)\simeq 0.741$ for $\alpha=100$, and
this value slightly increases when $\alpha$ decreases\footnote{
On large scales, $k<0.01~h/\mathrm{Mpc}$, 
the numerical ratio $B_{\fnl}^{nG1}/B_{\fnl}^{nG0}$ is constant, and
is equal to $3.15$, $3.06$, $3.00$, and $2.98$ for 
$\alpha=10$, $20$, $50$, and $100$, respectively.
}.
Therefore, for the squeezed triangle, we find a simple and illuminating result:
\begin{equation}
B_{\fnl}^{tot}(k_1,k_2,k_3)
\simeq
15
B_{\fnl}^{nG0}(k_1,k_2,k_3).
\label{eq:bfnl_tot_squeeze}
\end{equation}
This is an important result, showing that the statistical error on
$\fnl$ from the galaxy bispectrum will be smaller by at least a factor
of 15, compared to what was obtained in 
\citet{sefusatti/komatsu:2007}.

Figure~\ref{fig:z0f40} and the top panels of Figure~\ref{fig:mild_squeeze}
show various bispectrum terms in various
triangle configurations (see Fig.~\ref{fig:triangles} for the visual
representations of the triangles), evaluated at $z=0$.
As an example we use the following bias and non-Gaussianity parameters: 
${b_2}/{b_1}=0.5$, $\fnl=40$, and $\gnl=10^4$. 
The value of the linear bias, $b_1$, is irrelevant here as it does not
change the relative importance of  terms in
Eq.~(\ref{eq:bisp_nG2}), and thus we show the bispectrum terms divided by
$b_1^3$.

The message is quite simple: it is the squeezed configuration
that provides the best window into the primordial non-Gaussianity. The
other non-linear effects become more and more dominant as we move from
the squeezed to the equilateral, i.e., (a) to (e) in
Fig.~\ref{fig:triangles}. 
Even with this generous amount of non-Gaussian signals, $\fnl=40$ and
$\gnl=10^4$, 
only $\fnl$ term can be visible in the isosceles and equilateral
configurations on large scales.

For the the non-squeezed configurations, 
the $\fnl^2$ and $\gnl$ terms with the above chosen parameters are
comparable and the $\fnl$ term is order of magnitude 
greater than the $\fnl^2$ and $\gnl$
terms; however, the $\fnl^2$ term 
is the most dominant of all on large scales in the squeezed 
configuration ($\alpha>10$). 

We can understand these results analytically by comparing 
Eqs. (\ref{eq:bfnl_tot_squeeze}),
(\ref{eq:bfnl_squeeze}), 
(\ref{eq:bgnl_squeeze}), and (\ref{eq:bfnl2_squeeze}).
For the squeezed triangles with $k_1=k_2=\alpha k_3$ ($\alpha\gg1$)
and a scale-invariant spectrum, $P_\phi\propto k^{-3}$, we find
\begin{eqnarray}
\nonumber
\frac{\fnl B_{\fnl}^{tot}}{\fnl^2 B_{\fnl^2}^{nG}}
&\simeq&
\frac{15}{\fnl \alpha^2}\frac{\M0 k^2}{4\mathcal{I}_0 (\tilde{b}_2/b_1)}
\\
&\simeq&
0.0240
\left(\frac{100}{\alpha}\right)^2\frac{40}{\fnl}
\left(\frac{k}{0.01~h~\mathrm{Mpc}^{-1}}\right)^2,
\end{eqnarray}
\begin{eqnarray}
\nonumber
\frac{\gnl B_{\gnl}^{nG}}{\fnl^2 B_{\fnl^2}^{nG}}
&\simeq&
\frac{3}{2\alpha^2}\frac{\gnl}{\fnl^2}
\\
&\simeq&
0.000938
\left(\frac{100}{\alpha}\right)^2
\left(\frac{40}{\fnl}\right)^2
\frac{\gnl}{10^4},
\end{eqnarray}
\begin{eqnarray}
\nonumber
\frac{\fnl B_{\fnl}^{tot}}{\gnl B_{\gnl}^{nG}}
&\simeq&
15\frac{\fnl}{\gnl}
\frac{\M0 k^2}{6\mathcal{I}_0}
\\
&\simeq&
25.6\frac{\fnl}{40}
\frac{10^4}{\gnl}
\left(\frac{k}{0.01~h~\mathrm{Mpc}^{-1}}\right)^2.
\end{eqnarray}
These  estimates confirm that 
$B_{\fnl^2}$ dominates over $B_{\fnl}$ and $B_{\gnl}$ in the squeezed 
configurations on large scales, $k\lesssim 0.05~h~{\rm Mpc}^{-1}$ for 
$\alpha=100$, and $k\lesssim 0.03~h~{\rm Mpc}^{-1}$ for $\alpha=50$.
For $\alpha=10$, $\fnl^2$ term dominates only on the extremely large scales:
$k\lesssim 0.006~h~{\rm Mpc}^{-1}$.

Note that for a given configuration (for a given $\alpha$),
$B_{\fnl}^{nG}/B_{\fnl^2}^{nG}\propto k^2$ 
and 
$B_{\fnl}^{nG}/B_{\gnl}^{nG}\propto k^2$  while
$B_{\gnl}^{nG}/B_{\fnl^2}^{nG}$ is independent of $k$, which is
consistent with what we show in Fig.~\ref{fig:z0f40} on $k\lesssim
0.1~h~{\rm Mpc}^{-1}$.

In summary, the most unexpected and important results of our study are
as follows.
\begin{itemize}
\item
The terms that are linearly proportional to $\fnl$, derived in 
\citet{sefusatti/komatsu:2007}, receive additional contributions,
and are enhanced by a factor of $\sim 15$ for the squeezed triangles
(see Eq.~(\ref{eq:bfnl_tot_squeeze})).
\item
The $\fnl^2$ (or $\tau_{\rm NL}$) term 
actually dominates over the $\fnl$ term by a large factor for the squeezed
triangles (see the top-left panel of Fig.~\ref{fig:z0f40}).
\end{itemize}
This suggests that the galaxy bispectrum is more sensitive 
to $\fnl$ than previously recognized by \citet{sefusatti/komatsu:2007}, 
greatly enhancing our ability to detect the primordial non-Gaussianity
of local type. 
On very large scales, $k_1\ll 0.01~h~{\rm Mpc}^{-1}$, even the $\gnl$ term
(with $\gnl=10^4$)  dominates over the $\fnl$ term, giving us a hope
that perhaps we can obtain a meaningful limit on this term using the
galaxy bispectrum.

\subsection{Redshift Dependence}
The quantities we have calculated so far are evaluated at the present
epoch, $z=0$. 
At higher redshift,  each quantity needs to be scaled with
some powers of the linear growth factor $D(z)$,
which is normalized to 1 at the present epoch.
We find $P_R\propto D^2(z)$, $B_m^G\propto D^4(z)$, 
$B_{\fnl}^{nG0}\propto D^3(z)$,
$B_{m}^{nG}\propto D^3(z)$,
$B_{\fnl}^{nG1}\propto D^3(z)$,
$B_{\fnl^2}^{nG}\propto D^2(z)$, and $B_{\gnl}^{nG}\propto D^2(z)$.
Therefore, the final result for the halo bispectrum from the local type 
non-Gaussianity is 
\begin{widetext}
\begin{eqnarray}\label{eq:bhalo_final}
\nonumber
B_g(k_1,k_2,k_3,z)
&=&
b_1^3(z)D^4(z)
\left[
B_m^G(k_1,k_2,k_3)
+
\frac{b_2(z)}{b_1(z)}
\left\{P_R(k_1)P_R(k_2)+(2~\mbox{cyclic})\right\}
+
\frac{\fnl}{D(z)}
B_{\fnl}^{nG0}(k_1,k_2,k_3)
\right.
\\
\nonumber
&&+
\frac{\tilde{b}_2(z)}{b_1(z)}
\left\{
\frac{\fnl}{D(z)}
\left(
B_m^{nG}(k_1,k_2,k_3)
+
4\left(\Gr(k_1)+\Gr(k_2)+\Gr(k_3)\right)
B_{\fnl}^{nG0}(k_1,k_2,k_3)
+
B_{\fnl}^{nG1}(k_1,k_2,k_3)
\right)
\right.
\\
&&+
\left.\left.
\frac{\fnl^2}{D^2(z)}
B_{\fnl^2}^{nG}(k_1,k_2,k_3)
+
\frac{\gnl}{D^2(z)}
B_{\gnl}^{nG}(k_1,k_2,k_3)
\right\}\right],
\end{eqnarray}
\end{widetext}
where $B_m^G$, $P_R$, $B_{\fnl}^{nG0}$, $B_{m}^{nG}$, $B_{\fnl}^{nG1}$, $B_{\fnl^2}^{nG}$, and
$B_{\gnl}^{nG}$ are evaluated at $z=0$.

From equation (\ref{eq:bhalo_final}) it is clear that the contributions from 
non-Gaussian initial conditions become more and more important as we go 
to higher redshifts.
The new terms that we have derived in this paper,
the $B_{\fnl^2}^{nG}$ and $B_{\gnl}^{nG}$ terms,
are even more important than 
the term derived by \citet{sefusatti/komatsu:2007},  $B_{\fnl}^{nG0}$, 
This property makes high-redshift galaxy surveys particularly a 
powerful probe of primordial non-Gaussianity.

\begin{figure*}
\begin{center}
\rotatebox{90}{%
\includegraphics[width=12.5cm]{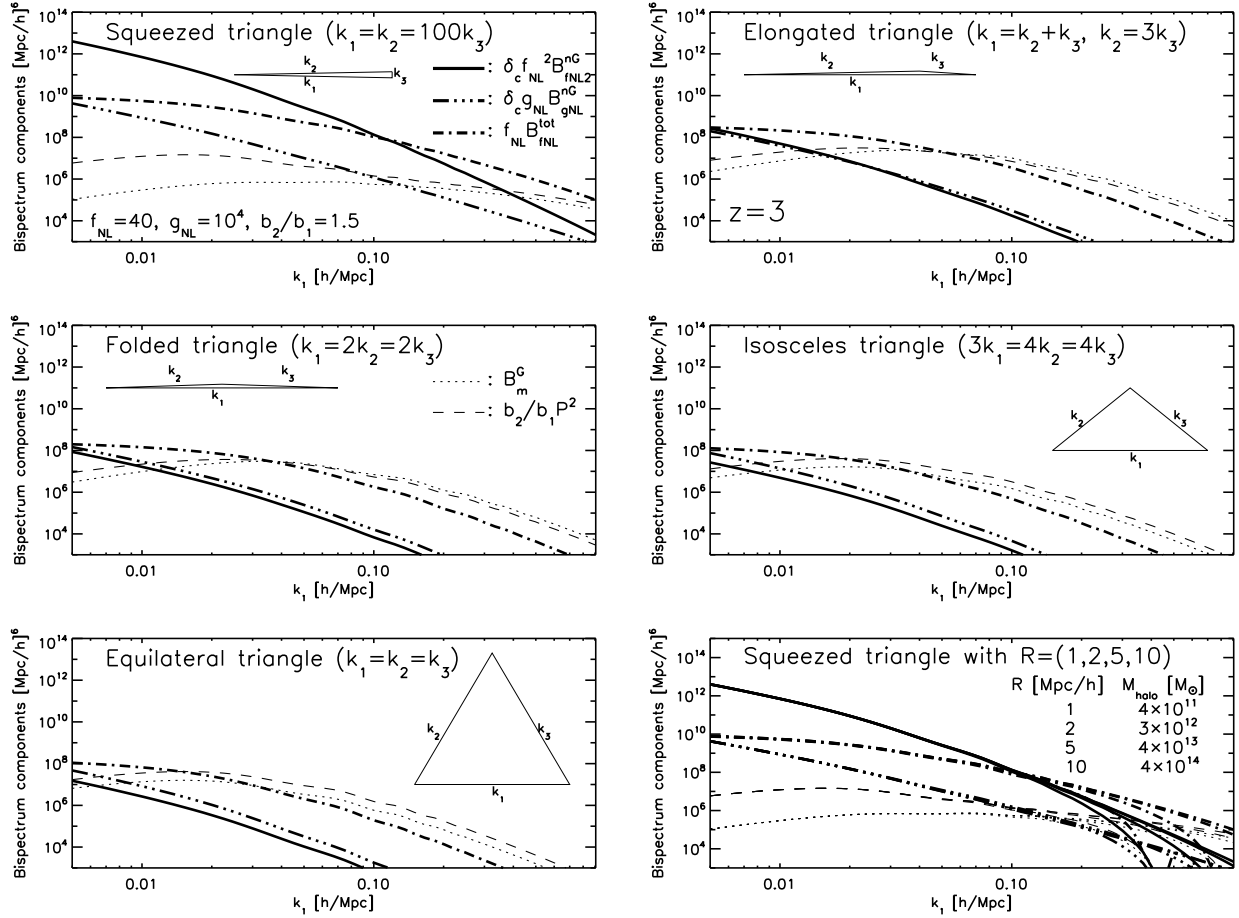}
}%
\caption{
 Same as Fig.~\ref{fig:z0f40}, but for $z=3$ and $b_2/b_1=1.5$. The
 non-Gaussianity parameters, 
 $\fnl=40$ and $\gnl=10^4$,  are the same as in
 Fig.~\ref{fig:z0f40}. 
         }%
\label{fig:z3f40}
\end{center}
\end{figure*}

Fig.~\ref{fig:z3f40} and the middle panel of Fig.~\ref{fig:mild_squeeze} show 
the bispectrum terms at $z=3$. 
Note that we use a larger value for the non-linear bias, $b_2/b_1=1.5$,
in accordance with a halo model \citep{cooray/sheth:2002}.
At this redshift, with $\fnl=40$ and $\gnl=10^4$, 
the $\gnl$ and $\fnl^2$
terms dominate over the non-linear effects 
also in the elongated, folded and isosceles configurations
at $k\lesssim 0.01~h~{\rm Mpc}^{-1}$,
as well as in the squeezed ones. 
The $\fnl$ terms dominate
over the non-linear effects on even smaller scales, and 
the importance of the $\fnl^2$
and $\gnl$ terms relative to the $\fnl$ term is greater, as expected
from their dependence on $D(z)$.

\section{Discussion and Conclusions}

Let us come back to this question, ``can we still use Sefusatti \&
Komatsu's equation, Eq.~(\ref{eq:SK07}), with $b_1$ replaced by the
scale-dependent bias, Eq.~(\ref{eq:bk})?''
The answer is clearly no: the primordial non-Gaussianity gives the
largest signal in the squeezed limit, whereas the non-linear
gravitational evolution and non-linear bias give the minimal signals in
the same limit.
This means that these effects are physically totally distinct, and thus
a mere scale-dependent rescaling of one effect does not give another.
Therefore, replacing $b_1$ in Eq.~(\ref{eq:SK07})
 with the scale-dependent bias in Eq.~(\ref{eq:bk})
results in an incorrect prediction.
For example, even though we have a term similar to that of the
scale-dependent bias, $B_m^{nG}$, in our final expression of the galaxy
bispectrum for the local-type primordial non-Gaussianity, 
there are many more terms that do not look like 
the scale-dependent bias that appears in the galaxy power spectrum.
Furthermore, $B_m^{nG}$ is by no means the most dominant term.

In this paper, we have derived a general expression for the bispectrum
of density peaks in the presence of primordial non-Gaussianity
(Eq.~(\ref{eq:bisp_nG})), using the MLB formula as well as using 
the local bias ansatz.
This result is general as long as we consider the bispectrum of 
 high density peaks, i.e., $\nu=\delta_c/\sigma_R\gg 1$, which is
 equivalent to highly biased galaxy populations, $b_1\gg 1$, 
on large scales in which the $n$-point correlation functions are much
smaller than unity. (This condition was necessary for us to 
Taylor expand the exponential in Eq.~(\ref{eq:P3_nG}).)

We have applied our formula to the local form of primordial
non-Gaussianity in Bardeen's curvature perturbations,
$\Phi=\phi+\fnl\phi^2+\gnl\phi^3$, and found new terms
that are proportional to 
$\fnl$, $\fnl^2$  and $\gnl$, which were absent in the formula derived by
\citet{sefusatti/komatsu:2007}.
We have examined the shape and scale dependence of these new terms as
well as those of the known 
terms, and found that the primordial non-Gaussianity contributions yield
the largest signals in the squeezed triangle configurations, where
the non-linear gravitational evolution and non-linear bias yield
the minimal signals. This is a good news: this property enables us
to distinguish the primordial and non-primordial effects easily.

The effects of primordial non-Gaussianity on the galaxy bispectrum are
more important in a high redshift universe, and thus high-redshift
galaxy surveys are particularly a potent probe of the physics of
inflation via measurements of primordial non-Gaussianity.

\begin{figure*}
\begin{center}
\rotatebox{90}{%
\includegraphics[width=12.5cm]{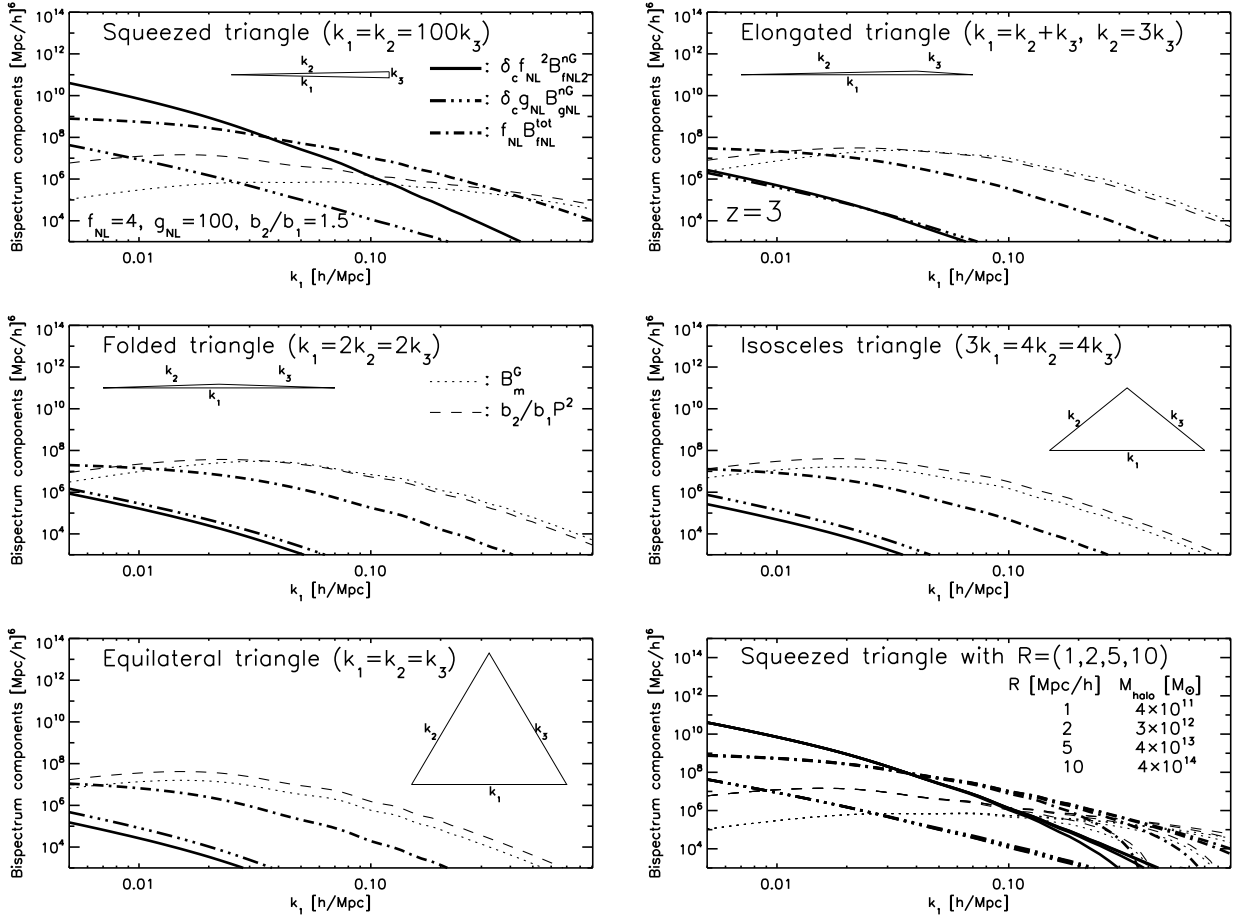}
}%
\caption{
 Same as Fig.~\ref{fig:z3f40}, but for smaller non-Gaussianity
 parameters,  $\fnl=4$ and $\gnl=100$.
         }%
\label{fig:z3f4}
\end{center}
\end{figure*}

The most significant conclusion of this paper is that, in the squeezed
configurations, the $\fnl^2$ term actually dominates over the 
$\fnl$ term by a large factor, and, on large scales,
newly derived $\fnl$ term dominates over the non linear terms for all 
configurations.
Because of this, the galaxy bispectrum should be more
sensitive to $\fnl$ than previously recognized: we have found a factor
of $\sim 15$ enhancement for the $\fnl$ term studied in
\citet{sefusatti/komatsu:2007}. 
In addition it is also
sensitive to a new term, $\gnl$.
Figure~\ref{fig:z3f4} and the bottom panel of Fig.~\ref{fig:mild_squeeze}
shows the bispectrum at $z=3$ with much
reduced primordial non-Gaussianity parameters, $\fnl=4$ and $\gnl=100$. 
In the squeezed configurations ($\alpha=100$), 
the $\fnl^2$ term is still well above
the usual terms from Gaussian initial conditions at $k\lesssim
0.1~h~{\rm Mpc}^{-1}$, the $\fnl$ term is above at 
$k\lesssim 0.4~{\rm Mpc}^{-1}$, and 
the $\gnl$ term is above at $k\lesssim 0.01~h~{\rm Mpc}^{-1}$.
Even with the milder squeezed limit for $\alpha=10$, the $\fnl$ term
still is above the Gaussian term at $k\lesssim 0.02~{\rm Mpc}^{-1}$.

The fact that the $\fnl^2$ term dominates in the squeezed limit 
is particularly interesting, as it provides us with the unique
window into the physics of inflation in the following way.
Recently, a number of groups
\citep[e.g.,][]{boubekeur/lyth:2006,huang/shiu:2006,byrnes/sasaki/wands:2006,buchbinder/khoury/ovrut:2008} 
have shown that the primordial
trispectrum can in general be written as 
\begin{eqnarray}
\nonumber
&&T_\Phi(\mathbi{k}_1,\mathbi{k}_2,\mathbi{k}_3,\mathbi{k}_4)
\\
\nonumber
&
=
&
6\gnl 
\left[
P_\phi(k_1) P_\phi(k_2) P_\phi(k_3) + (3~\mathrm{cyclic})
\right]
+\frac{25}{18}\tau_\mathrm{NL}
\\
&&\times
\left[
P_\phi(k_1) P_\phi(k_2)
\left\{
 P_\phi(k_{13})
+ P_\phi(k_{14})
\right\}
+
(11~\mathrm{cyclic})
\right],
\end{eqnarray}
instead of Eq.~(\ref{eq:tphi}). Different models of the early universe
predict different relations between $\tau_\mathrm{NL}$ and $\fnl$. 
Therefore, separately detecting the $\tau_{\rm NL}$ (i.e., $\fnl^2$) and
$\fnl$ terms can be a powerful tool for constraining the model of the
early universe. 

How well one can constrain these
parameters with the planned future high-redshift galaxy surveys will be
presented in a forthcoming paper.

\acknowledgments
E.K. would like to thank Kazuya Koyama for useful discussions which led
to \S~2.2, 
and the organizers for the ``Focus Week on Non-Gaussianities in the Sky,'' 
where he received useful comments on the draft of this paper.
We also would like to thank Emiliano Sefusatti for useful comments. 
While we are adding \S~2.2 and the additional trispectrum terms,
$T_R^{1112}$, to the first version of the preprint posted to arXiv:0904.0497,
\citet{sefusatti:prep}, which performed similar calculations, appeared
on arXiv:0905.0717. Our alternative derivation  given in
\S~2.2 and additional calculations on $T_R^{1112}$ are independent of,
and agree with, those given in \citet{sefusatti:prep}.
This material is based in part upon work
supported by the Texas Advanced Research Program under
Grant No. 003658-0005-2006, by NASA grants NNX08AM29G
and NNX08AL43G, and by NSF grants AST-0807649 and PHY-0758153.
E.K. acknowledges support from an Alfred P. Sloan Fellowship.
D.J. acknowledges support from a  Wendell Gordon Endowed Graduate 
Fellowship of the University of Texas at Austin.

\begin{appendix}
\section{A. Integration of $T_R^{1112}$}\label{sec:appA}
In the standard perturbation theory, the four-point correlator contained
in the definition of
$T_R^{1112}$ (see Eq.~(\ref{eq:def_Tijkl})) is given by\footnote{\citet{sefusatti:prep} also derived and
studied this term independently.}
\begin{eqnarray}
\label{eq:Tr1112_1st}
\langle
\delta^{(1)}(\mathbi{k}_1)
\delta^{(1)}(\mathbi{k}_2)
\delta^{(1)}(\mathbi{k}_3)
\delta^{(2)}(\mathbi{k}_4)
\rangle
=
\int\frac{d^3q}{(2\pi)^3}
F_2^{(s)}(\mathbi{q},\mathbi{k}_4-\mathbi{q})
\langle
\delta^{(1)}(\mathbi{k}_1)
\delta^{(1)}(\mathbi{k}_2)
\delta^{(1)}(\mathbi{k}_3)
\delta^{(1)}(\mathbi{k}_4-\mathbi{q})
\delta^{(1)}(\mathbi{q})
\rangle.
\end{eqnarray}
For non-Gaussian density fields, the leading order of Eq.~(\ref{eq:Tr1112_1st}) 
contains the ensemble average of products of six Gaussian variables,
$\phi$, which gives products of three power spectra, $P_\phi$.
We find
\begin{eqnarray}
\nonumber
&&
\int\frac{d^3q}{(2\pi)^3}
F_2^{(s)}(\mathbi{q},\mathbi{k}_4-\mathbi{q})
\langle
\delta^{(1)}(\mathbi{k}_1)
\delta^{(1)}(\mathbi{k}_2)
\delta^{(1)}(\mathbi{k}_3)
\delta^{(1)}(\mathbi{k}_4-\mathbi{q})
\delta^{(1)}(\mathbi{q})
\rangle
\\
\nonumber
&=&
\int\frac{d^3q}{(2\pi)^3}
F_2^{(s)}(\mathbi{q},\mathbi{k}_4-\mathbi{q})
\langle
\delta^{(1)}(\mathbi{k}_1)
\delta^{(1)}(\mathbi{k}_2)
\rangle
\langle
\delta^{(1)}(\mathbi{k}_3)
\delta^{(1)}(\mathbi{k}_4-\mathbi{q})
\delta^{(1)}(\mathbi{q})
\rangle
+(\mathrm{cyclic}~123)
\\
\nonumber
&&+
2 \int\frac{d^3q}{(2\pi)^3}
F_2^{(s)}(\mathbi{q},\mathbi{k}_4-\mathbi{q})
\langle
\delta^{(1)}(\mathbi{k}_1)
\delta^{(1)}(\mathbi{q})
\rangle
\langle
\delta^{(1)}(\mathbi{k}_2)
\delta^{(1)}(\mathbi{k}_3)
\delta^{(1)}(\mathbi{k}_4-\mathbi{q})
\rangle
+(\mathrm{cyclic}~123)
\\
\nonumber
&=&
(2\pi)^3 
\biggl[
2\fnl P_m(k_1) \Mm(k_3)
\int d^3q
\Mm(q)\Mm(|\mathbi{k}_4-\mathbi{q}|)
P_\phi(q)
\left\{
P_\phi(|\mathbi{k}_4-\mathbi{q}|)
+2P_\phi(k_3)
\right\}
F_2^{(s)}(\mathbi{q},\mathbi{k}_4-\mathbi{q})
\delta^D(\mathbi{k}_{12})
\\
\nonumber
&&
+
4\fnl \Mm(k_2)\Mm(k_3)\Mm(k_{14})P_m(k_1)
F_2^{(s)}(-\mathbi{k}_1,\mathbi{k}_{14})
\\
&&\times
\left\{
P_\phi(k_2)P_\phi(k_3)
+
P_\phi(k_2)P_\phi(k_{14})
+
P_\phi(k_3)P_\phi(k_{14})
\right\}
+(\mathrm{cyclic}~123)
\biggl]
\delta^D(\mathbi{k}_{1234}).
\end{eqnarray}
Therefore, $T_R^{1112}$ is given by
\begin{eqnarray}
\nonumber
&&
T_R^{1112}(\mathbi{k}_1,\mathbi{k}_2,\mathbi{k}_3,\mathbi{k}_4)
\\
&=&
\nonumber
{\Wr(k_1)\Wr(k_2)\Wr(k_3)\Wr(k_4)}
\\
\nonumber
&&\times
\biggl[
2\fnl P_m(k_1) \Mm(k_3)
\int d^3q
\Mm(q)\Mm(|\mathbi{k}_4-\mathbi{q}|)
P_\phi(q)
\left\{
P_\phi(|\mathbi{k}_4-\mathbi{q}|)
+2P_\phi(k_3)
\right\}
F_2^{(s)}(\mathbi{q},\mathbi{k}_4-\mathbi{q})
\delta^D(\mathbi{k}_{12})
\\
\nonumber
&&
+
4\fnl \Mm(k_2)\Mm(k_3)\Mm(k_{14})P_m(k_1)
F_2^{(s)}(-\mathbi{k}_1,\mathbi{k}_{14})
\\
&&\times
\left\{
P_\phi(k_2)P_\phi(k_3)
+
P_\phi(k_2)P_\phi(k_{14})
+
P_\phi(k_3)P_\phi(k_{14})
\right\}
+(\mathrm{cyclic}~123)
\biggl],
\label{eq:Tr1112}
\end{eqnarray}
where $\Mm(k)\equiv\Mr(k)/\Wr(k)$,
$\mathbi{k}_{ij}=\mathbi{k}_i+\mathbi{k}_j$, and 
$(\mathrm{cyclic}~123)$ denotes that the cyclic changes among
$(\mathbi{k}_1$, $\mathbi{k}_2$, $\mathbi{k}_3)$.
We calculate the sum of $\{1112\}$ terms in Eq.~(\ref{eq:Tr_separate}) by 
integrating Eq.~(\ref{eq:Tr1112}):
\begin{eqnarray}
\nonumber
&&\int\frac{d^3q}{(2\pi)^3}
T_R^{(2)}(\mathbi{q},\mathbi{k}_1-\mathbi{q},\mathbi{k}_2,\mathbi{k}_3)
\\
\nonumber
=&&
8\fnl\Wr(k_2)\Wr(k_3)
\Mm(k_1)\Mm(k_2)\Mm(k_{3})
\left\{
P_\phi(k_2)P_\phi(k_3)
+
P_\phi(k_2)P_\phi(k_{1})
+
P_\phi(k_3)P_\phi(k_{1})
\right\}
\\
\nonumber
&&\times
\int\frac{d^3q}{(2\pi)^3}
\Wr(|\mathbi{k}_1-\mathbi{q}|)
\Wr(q)
P_m(q)
F_2^{(s)}(-\mathbi{q},\mathbi{k}_{1})
\\
\nonumber
&+&
4\fnl\Wr(k_2)\Wr(k_3)
\Mm(k_{1})
\left[
P_m(k_2)
F_2^{(s)}(\mathbi{k}_2,\mathbi{k}_{1})
+
P_m(k_3)
F_2^{(s)}(\mathbi{k}_3,\mathbi{k}_{1})
\right]
\\
\nonumber
&&\times
\int\frac{d^3q}{(2\pi)^3}
\Wr(|\mathbi{k}_1-\mathbi{q}|)
\Wr(q)
\Mm(q)\Mm(|\mathbi{k}_1-\mathbi{q}|)
\left\{
P_\phi(q)P_\phi(|\mathbi{k}_1-\mathbi{q}|)
+
2P_\phi(q)P_\phi(k_{1})
\right\}
\\
\nonumber
&+&
8\fnl\Wr(k_2)\Wr(k_3)
\Mm(k_3)P_m(k_2)
\int\frac{d^3q}{(2\pi)^3}
\Wr(|\mathbi{k}_1-\mathbi{q}|)
\Wr(q)
\Mm(|\mathbi{k}_1-\mathbi{q}|)\Mm(|\mathbi{k}_2+\mathbi{q}|)
F_2^{(s)}(-\mathbi{k}_2,\mathbi{k}_{2}+\mathbi{q})
\\
\nonumber
&&\times
\left\{
P_\phi(k_3)P_\phi(|\mathbi{k}_1-\mathbi{q}|)
+
P_\phi(k_3)P_\phi(|\mathbi{k}_{2}+\mathbi{q}|)
+
P_\phi(|\mathbi{k}_1-\mathbi{q}|)P_\phi(|\mathbi{k}_{2}+\mathbi{q}|)
\right\}
+(k_2 \leftrightarrow k_3)
\\
\nonumber
&+&
8\fnl\Wr(k_2)\Wr(k_3)
\Mm(k_3)
\int\frac{d^3q}{(2\pi)^3}
\Wr(|\mathbi{k}_1-\mathbi{q}|)
\Wr(q)
\Mm(|\mathbi{k}_1-\mathbi{q}|)
\Mm(|\mathbi{k}_{2}+\mathbi{q}|)P_m(q)
F_2^{(s)}(-\mathbi{q},\mathbi{k}_{2}+\mathbi{q})
\\
\nonumber
&&\times
\left\{
P_\phi(|\mathbi{k}_1-\mathbi{q}|)P_\phi(k_3)
+
P_\phi(|\mathbi{k}_1-\mathbi{q}|)P_\phi(|\mathbi{k}_{2}+\mathbi{q}|)
+
P_\phi(k_3)P_\phi(|\mathbi{k}_{2}+\mathbi{q}|)
\right\}
+(k_2 \leftrightarrow k_3)
\\
\nonumber
&+&
8\fnl 
\left(\Wr(k_2)\Wr(k_3)\right)^2
P_m(k_3) \Mm(k_2)
\\
&&\times
\int \frac{d^3q}{(2\pi)^3}
\Mm(q)\Mm(|\mathbi{k}_2-\mathbi{q}|)
P_\phi(q)
\left\{
P_\phi(|\mathbi{k}_2-\mathbi{q}|)
+2P_\phi(k_2)
\right\}
F_2^{(s)}(\mathbi{q},\mathbi{k}_2-\mathbi{q})
+(k_2 \leftrightarrow k_3).
\label{eq:int_Tr1112}
\end{eqnarray}

\section{B. Summary of Equations}\label{sec:appB}
As various terms contributing to the galaxy bispectrum are scattered
over various places in  the paper, 
we collect them together in this Appendix. For galaxies of size $R$ 
(or mass $M=(4\pi/3)R^3\bar{\rho}_m$, where $\bar{\rho}_m$ is the 
cosmic mean matter density), the galaxy bispectrum at redshift $z$ is
given by
\begin{eqnarray}
\nonumber
B_g(k_1,k_2,k_3,z)
&=&
b_1^3(z)D^4(z)
\left[
B_m^G(k_1,k_2,k_3)
+
\frac{b_2(z)}{b_1(z)}
\left\{P_R(k_1)P_R(k_2)+(2~\mbox{cyclic})\right\}
+
\frac{\fnl}{D(z)}
B_{\fnl}^{nG0}(k_1,k_2,k_3)
\right.
\\
\nonumber
&&+
\frac{\tilde{b}_2(z)}{b_1(z)}
\left\{
\frac{\fnl}{D(z)}
\left(
B_m^{nG}(k_1,k_2,k_3)
+
4\left(\Gr(k_1)+\Gr(k_2)+\Gr(k_3)\right)
B_{\fnl}^{nG0}(k_1,k_2,k_3)
+
B_{\fnl}^{nG1}(k_1,k_2,k_3)
\right)
\right.
\\
&&+
\left.\left.
\frac{\fnl^2}{D^2(z)}
B_{\fnl^2}^{nG}(k_1,k_2,k_3)
+
\frac{\gnl}{D^2(z)}
B_{\gnl}^{nG}(k_1,k_2,k_3)
\right\}\right],
\label{eq:Bg_summary}
\end{eqnarray}
where $b_1(z)$ and $b_2(z)$ are the linear and non-linear bias parameters, 
respectively. As we mentioned in \S 2, 
$\tilde{b}_2(z)/b_1(z)$ would be equal to 
$\delta_c$ within the context of the MLB formalism, but the precise
value has to be measured from N-body simulations.

Note that the redshift evolution of each term in explicitly given by
the powers of the linear growth factor $D(z)$, and various contributions,
$B_m^G$, $P_R$, $B_m^{nG}$, 
$B_{\fnl}^{nG0}$, $B_{\fnl}^{nG1}$, $B_{\fnl^2}^{nG}$, and $B_{\gnl}^{nG}$, are 
evaluated at $z=0$ with
\begin{eqnarray}
B_m^G(k_1,k_2,k_3)
&=&
2F_2^{(s)}(\mathbi{k}_1,\mathbi{k}_2)
\Wr(k_1)\Wr(k_2)\Wr(k_3)
P_m(k_1)P_m(k_2)+(2~\mbox{cyclic})
\\
B_{\fnl}^{nG0}(k_1,k_2,k_3)
&=&
2
\frac{P_R(k_1)}{\Mr(k_1)}
\frac{P_R(k_2)}{\Mr(k_2)}\Mr(k_3) + (2~\mbox{cyclic})
\\
B_{m}^{nG}(k_1,k_2,k_3)
&=&
4\Wr(k_1)\Wr(k_2)\Wr(k_3)
\left[
\frac{\Fr(k_1)}{\Mr(k_1)}
+
\frac{\Fr(k_2)}{\Mr(k_2)}
\right] P_m(k_1)P_m(k_2)F_2^{(s)}(\mathbi{k}_1,\mathbi{k}_2)
+(2~\mbox{cyclic})
\\
\nonumber
B_{\fnl}^{nG1}(k_1,k_2,k_3)
&\approx&
\frac{1}{2\sigma_R^2}
\biggl[
8\Wr(k_2)\Wr(k_3)
P_m(k_2)\Mm(k_3)P_\phi(k_3)
\int \frac{d^3q}{(2\pi)^3} 
\Wr(|\mathbi{k}_1-\mathbi{q}|)
\Wr(q)
\Mm(|\mathbi{k}_1-\mathbi{q}|)
\Mm(|\mathbi{k}_2+\mathbi{q}|)
\\
&&\times
\left[
P_\phi(|\mathbi{k}_2+\mathbi{q}|) + 
P_\phi(|\mathbi{k}_1-\mathbi{q}|)
\right]
F_2^{(s)}(-\mathbi{k}_2,\mathbi{k}_2+\mathbi{q})
+(5~\mathrm{permutation})\bigg]
\\
\nonumber
B_{\fnl^2}^{nG}(k_1,k_2,k_3)
&\approx &
\frac{1}{2\sigma_R^2}\biggl[
8\Mr(k_2)\Mr(k_3)
P_\phi(k_1)
\left[
P_\phi(k_2)+P_\phi(k_3)
\right]
\int \frac{d^3q}{(2\pi)^3}
\Mr(q)
\Mr(|\mathbi{k}_1-\mathbi{q}|)
P_\phi(q)
+
(2~\mathrm{cyclic})
\\
\nonumber
&&+
4\Mr(k_2)\Mr(k_3)P_\phi(k_2)P_\phi(k_3)
\int \frac{d^3q}{(2\pi)^3}
\Mr(q)
\Mr(|\mathbi{k}_1-\mathbi{q}|)
\\
& &\times
\left[
P_\phi(|\mathbi{k}_2+\mathbi{q}|)
+
P_\phi(|\mathbi{k}_3+\mathbi{q}|)
\right]
+
(2~\mathrm{cyclic})\biggr]
\\
\nonumber
B_{\gnl}^{nG}(k_1,k_2,k_3)
&=&
\frac{1}{2\sigma_R^2}\biggl[
6
\Mr(k_2)\Mr(k_3)\left[
P_\phi(k_2)+P_\phi(k_3)
\right]
\int \frac{d^3q}{(2\pi)^3}
\Mr(q)\Mr(|\mathbi{k}_1-\mathbi{q}|)
P_\phi(q)
P_\phi(|\mathbi{k}_1-\mathbi{q}|)+(2~\mathrm{cyclic})
 \\ 
&+&
12
\Mr(k_2)\Mr(k_3)P_\phi(k_2)P_\phi(k_3)
\int \frac{d^3q}{(2\pi)^3}
\Mr(q)\Mr(|\mathbi{k}_1-\mathbi{q}|)
P_\phi(q)
+(2~\mathrm{cyclic})\biggl].
\end{eqnarray}
Note that we show only dominant terms for $B_{\fnl}^{nG1}$ and 
$B_{\fnl^2}^{nG}$ on large scales. One can find the exact definitions 
in Eq.~(\ref{eq:bfnl2*}) and Eq.~(\ref{eq:BfnlnG1*}).
Finally, $\Fr(k)$ and $\Gr(k)$ are defined as follows.
\begin{eqnarray}
\Fr(k)
&\equiv&
\frac{1}{2\sigma_R^2}
\int\frac{d^3q}{(2\pi)^3}
P_\phi(q)
\Mr(q)
\Mr(|\mathbi{k}-\mathbi{q}|)
\left[
\frac{P_\phi(|\mathbi{k}-\mathbi{q}|)}{P_\phi(k)} +2
\right]
\\
\Gr(k)
&\equiv&
\frac{1}{2\sigma_R^2}
\int \frac{d^3q}{(2\pi)^3} 
\frac
{\Wr(q)\Wr(|\mathbi{k}-\mathbi{q}|)}{\Wr(k)}
P_m(q)F_2^{(s)}(\mathbi{k},-\mathbi{q})
\end{eqnarray}

\end{appendix}

\end{document}